\documentclass[usenatbib]{mn2e}
\usepackage{lipsum}
\usepackage{float}
 \usepackage{graphicx}
\usepackage{epstopdf}
\usepackage[caption=false]{subfig}
\usepackage{cprotect}
\usepackage{url}
\usepackage{amsmath} 
\usepackage{graphicx}
\usepackage{amssymb}
\usepackage{amstext}
\usepackage{floatpag}
\usepackage{float}
\usepackage[section]{placeins}
\usepackage{relsize}
\usepackage{amsmath}
\usepackage{url}
\usepackage{longtable}
\usepackage{xcolor}
\usepackage{color}
\usepackage{supertabular}
\usepackage[normalem]{ulem} 

\newcommand{\prd}{Physics Review D}

\newcommand{\apj}{ApJ}
\newcommand{\mnras}{MNRAS}
\newcommand{\aap}{A\&A}

\newcommand{\uflux}{$ \mathrm{erg}\, \mathrm{sec}^{-1}\, \mathrm{cm}^{-2}$}

\def \prd {PRD}
\def \apj {ApJ}
\def \apjs {ApJS}
\def \apjl {ApJL}
\def \mnras {MNRAS}
\def \aap {A\&A}
\def \aj {AJ}
\def \araa {ARAA}
\def \pasj {PASJ}
\def \pasa {PASA}

\def \nat {Nature}

\def \nar {New. A. Rev.}

\def \cjaa {Chin. J. Astron. Astrophys.}

\newcommand{\degree}{$^{\circ}$}

\title[Fast response EM follow-up prospects from low latency GW triggers]{Capturing the electromagnetic counterparts of binary neutron star mergers through low latency gravitational wave triggers}

\author[chu et al.]{ Q. Chu\thanks{E-mail:qi.chu@uwa.edu.au}$^{1}$, E. J. Howell\thanks{E-mail:eric.howell@uwa.edu.au}$^{1}$, A. Rowlinson \thanks{E-mail:b.a.rowlinson@uva.nl}$^{2,3,4}$, H. Gao$^{5}$, B. Zhang$^{6}$, S. J. Tingay$^{7,8}$ \and M. Bo\"er$^{9}$ and L. Wen$^{1}$\thanks{E-mail:linqing.wen@uwa.edu.au}  \\
$^{1}$School of Physics, University of Western Australia, Crawley WA 6009, Australia\\
$^{2}$CSIRO Astronomy and Space Science, Sydney, Australia \\
$^{3}$Anton Pannekoek Institute, University of Amsterdam, Postbus 94249, 1090 GE, Amsterdam\\
$^{4}$The Netherlands Netherlands Institute for Radio Astronomy (ASTRON), PO Box 2, 7990 AA Dwingeloo, The Netherlands\\
$^{5}$Department of Astronomy, Beijing Normal University, Beijing 100875, China\\
$^{6}$Department of Physics and Astronomy, University of Nevada Las Vegas, NV 89154, USA\\
$^{7}$International Centre for Radio Astronomy Research, Curtin University, Perth, WA 6845, Australia\\
$^{8}$ARC Centre of Excellence for All-sky Astrophysics (CAASTRO)\\
$^{9}$CNRS - ARTEMIS, boulevard de l'Observatoire, CS 34229, 06304 Nice Cedex 04, France}

\begin{document}
\pagerange{\pageref{firstpage}--\pageref{lastpage}} \pubyear{2002}
\maketitle

\label{firstpage}
\begin{abstract}
{We investigate the prospects for joint low-latency gravitational wave (GW) detection and prompt electromagnetic (EM) follow-up observations of coalescing binary neutron stars (BNSs). For BNS mergers associated with short duration gamma-ray bursts (SGRBs), we for the first time evaluate the feasibility of rapid EM follow-ups to capture the prompt emission, early engine activity or reveal any potential by-products such as magnetars or fast radio bursts. To achieve our goal, we first simulate a population of coalescing BNSs using realistic distributions of source parameters and estimate the detectability and localisation efficiency at different times before merger. We then use a selection of facilities with GW follow-up agreements in place, from low-frequency radio to high energy $\gamma$-ray to assess the prospects of prompt follow-up. We quantify our assessment using observational SGRB flux data extrapolated to be within the horizon distances of the advanced GW interferometric detectors LIGO and Virgo and to the prompt phase immediately following the binary merger. Our results illustrate that while challenging, breakthrough multi-messenger science is possible with EM follow-up facilities with fast responses and wide fields-of-view. We demonstrate that the opportunity to catch the prompt stage ($<$ 5s) of SGRBs, can be enhanced by speeding up the detection pipelines of both GW observatories and EM follow-up facilities. We further show that the addition of an Australian instrument to the optimal detector network could possibly improve the angular resolution by a factor of two and thereby contribute significantly to GW-EM multi-messenger astronomy.
}
\end{abstract}


\graphicspath{{figs/}{./}}
\section{Introduction}
\label{intro}
The LIGO detection of GW150914 from the inspiral and merger of a pair of black holes has ignited a new era of astronomy \citep{2016PhRvL.116f1102A}. To fully exploit this new frontier it has been expected that triggered follow-ups by facilities operating outside the GW spectrum will become common place. During this new multi-messenger era, prompt electromagnetic follow-ups from low-latency GW triggers can provide a pan-spectral pathway of understanding into the inner engines and underlying mechanisms behind a multitude of GW sources.\\
The Advanced LIGO \citep[aLIGO;][]{aLIGO_2015} detector launched its first science operations at sensitivities that would allow the detection of coalescing systems of binary neutron stars (BNSs) in the range of 40-80\,Mpc\footnote{This range refers to the detectable range averaged over all sky locations and source orientations for BNS systems with individual masses of 1.4 $M_{\odot}$. It is approximately 1/2.26 of the horizon distance - the range for optimally oriented and located sources.}; aLIGO will be followed by Advanced Virgo \citep[AdV;][]{AdV_2015} later this year (2016). By 2019 aLIGO is expected to reach design sensitivity, with a BNS detection range of 200Mpc. AdV is expected to reach this milestone by 2018 with BNS observations out to a distance of 130Mpc~\citep{AdV_2015}. Other second-generation detectors are also planned or undergoing construction. KAGRA, a Japanese instrument, is envisioned to begin operation in 2017~\citep{kagra} and fulfill its baseline design around 2018-2019~\citep{kagra_2012}. LIGO-India~\citep{ligo_localization} has obtained approval and is expected to be operational from 2020, reaching a design sensitivity at the same level as aLIGO no earlier than 2022~\citep{ligo_localization}.

Coalescing systems of BNSs are promising sources of GWs due to the fact that the inspiral phase is well modelled by the post-Newtonian (PN) approximation. The relatively low masses of BNS systems mean that these events will span most of the frequency band of aLIGO detectors.
At design sensitivity, aLIGO is predicted to make 0.4 to 400 BNS detections per year according to current rate predictions \citep{Abadie2010CQGra,CowardHowellPiran_2012}.
The coalescence rate of of binary systems comprising of neutron stars and black holes (NS/BH) and binary black holes (BBH) are less confidently predicted due to a sparsity of observational data. Studies have predicted yearly detection rates of order 0.2 -- 300 detections for NS/BH events and  0.4 -- 1000 for BBH events~\citep{Abadie2010}. For systems including at least one BH the detection of GW150914 is important as it proved that BH birth kicks do not always break up the evolutionary formation channels for these sources \citep{2016ApJ...818L..22A}.

\begin{figure*}
\includegraphics[scale = 0.1]{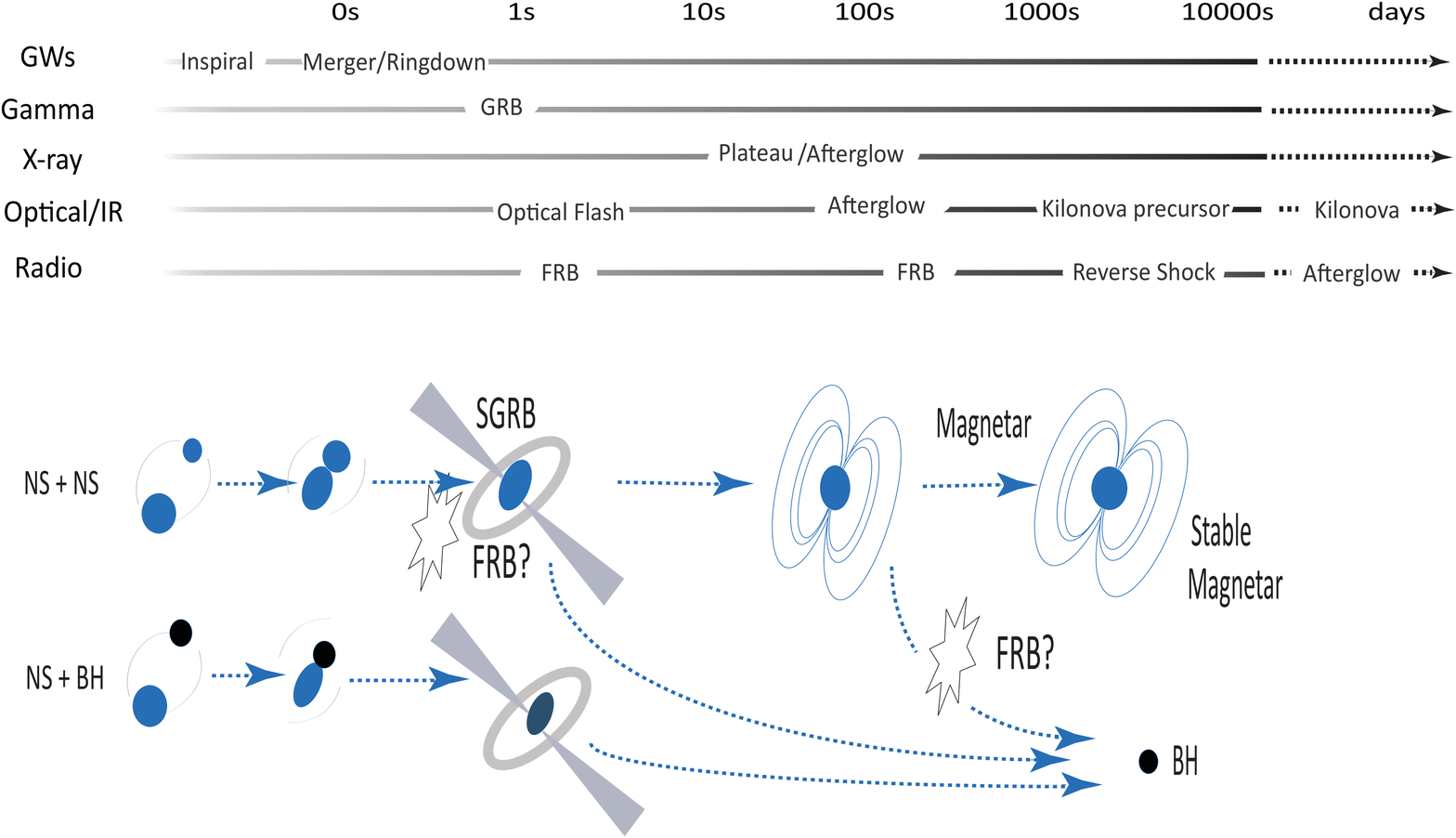}\\
\caption{A cartoon illustrating some of the possible multi-messenger pathways and end products of coalescing systems of NSs and BHs in different energy bands. We also show the approximate timescales from GW merger. Short duration gamma ray bursts (SGRBs) are strongly linked
with the merger of of compact objects \citep{Paczynski1986ApJ,Eichler1989Natur,Narayan_1992ApJ,Rezzolla_2011ApJ,gehrels05,Berger2005Natur, Bloom2006ApJ} and could be accompanied by an optical flash or a fast radio burst \citep[FRB][]{Totani2013PASJ}. If a stable or supra-massive magnetar is formed, the long lived X-ray plateaus observed in many SGRBs could be the signature of direct dissipation of magnetar wind energy
\citep{Rowlinson2010MNRAS,Rowlinson2013MNRAS,Zhang2013ApJ}; the collapse of a merger product to a black hole could also produce an FRB \citep{Zhang2014ApJ,Falcke_Rezzolla2014AA}. At later times predictions for an
optical/IR kilonova \citep{lipaczynski98,metzger10,barnes13} have gained support though recent observations \citep{Berger2013ApJ,Tanvir2013Natur}. For a relatively small opacity, a kilonova precursor may appear early enough which would benefit from the low latency observations \citep{Metzger_2014}.}
  \label{fig_mm_scenarios}
\end{figure*}

Multi-messenger astronomy is one of the highly prioritized research areas in the GW community \citep{Howell2015PASA}. Complementary observations from electromagnetic (EM), neutrino or high energy particle facilities would not only improve the confidence of a GW detection, but also maximise the science return by providing a wealth of additional astrophysical information. A number of plausible EM emission mechanisms and end products that could be produced during the inspiral and merger phase of coalescing compact objects have been proposed. We summarise some of the popular scenarios in Figure~\ref{fig_mm_scenarios};  predictions are highly uncertain as illustrated by the various pathways and end products and could involve combinations of BHs and/or magnetar formation or Fast Radio Bursts (FRBs). The anticipated onset time, duration, and observable wavelengths shown suggest that transient EM emission could possibly occur within a few seconds after the binary merger \citep{MetzgerBerger2012ApJ,Centrella2012IAUS}. To catch these events, generating GW event triggers without delay (thus low-latency) and obtaining prompt EM follow-ups are crucial. Achieving joint GW-EM observations of these events could reveal the major processes and interactions at play during the binary merger process (see detailed discussion in section~\ref{sec_astro}).

A significant effort from both GW and EM astronomy community has been undertaken to achieve prompt GW-triggered EM followup observations \citep{abadie11,Shawhan2012SPIE}, though as yet, no joint GW-EM detections have been made. Recent developments in the areas of astronomical and computing technologies in the following three fronts will further improve the chance of catching EM transients during the binary coalescence:
\begin{enumerate}
\item[(i)] Advances in low-latency GW search pipelines that are capable of generating event triggers within minutes upon the arrival of a detectable signal.
\item[(ii)] Localisations are likely to improve as more GW detectors are constructed.
\item[(iii)] Development of wide-field EM telescopes capable of rapid response within tens of seconds and with fields of view (FoV) comparable to the error area of GW source sky direction (typically tens of square degrees for the aLIGO/AdV network).
\end{enumerate}

The first point has been studied since the initial LIGO era and a number of low-latency GW trigger-generation pipelines have been proposed and tested ~\citep{first_low_latency_inspiral, mbta,  cannon12, jing,spiir}; the prospects of rapid alerts from GW events are discussed in ~\citet{Shawhan2012SPIE}.

The benefits of a larger network given in the second point have prompted studies of source localization by advanced networks by a number of authors ~\citep[see the studies of:][]{ligo_localization,fairhurst11,nissanke12,frequentist_network,schutz11,chuqi12}. \citet{wen_fan, fairhurst09,wen10, frequentist_single,abadie,samaya,klimenko} have investigated key aspects of source localisation and estimation of localisation accuracy. For low-latency detections, studies of the early localization of BNS sources when their GWs are detected 10s of seconds before the merger can be found in \citet{manzotti12,cannon12}.

A recent study of the sky localization ~\citep{first2years} for the early detector network expected in 2016-17 indicate that the error in GW localization will be of order 200 deg$^2$. Methods of identifying GW sources from a large sky error regions will therefore have great importance and have been proposed by \citet{Nissanke2013ApJ,identify_host_galaxies,cowperthwaite2015}.

To combat the large error regions the range of wide-field EM instruments that are in operation or are being developed across the entire range of the EM spectrum will be important. Additionally, the possibility of capturing the early EM emissions can be provided through low-latency alerts; this will require fast response EM instruments. A range of such instruments from high to low energy will be fully discussed in section 4.

This paper will investigate the types of breakthrough science that will be possible through the low-latency detection and localisation of BNS GW events, in particular, how these will enable prompt GW-EM observations of BNS coalescence in the era of advanced detectors.  While the science of GW-EM astronomy has been widely investigated previously for longer time scale EM emissions \cite[e.g.]{Howell2015PASA,MetzgerBerger2012ApJ} the feasibility to capture prompt and early EM emissions from binary mergers has not been studied before. To examine the possible EM responses, we use a comprehensive simulation that extends on previous investigations of the early detection and localization of compact binary coalescence (CBC) sources by extensively sampling a range of parameters of the NS waveform and sky localisation. We produce statistics showing the expected number of detections and their sky localisation error regions at different times before merger. Based on these results, we for the first time extrapolate the temporally complete EM data currently available for SGRBs to investigate the capacity of EM instruments for short time scale EM follow-up observations through the $\gamma$-ray, X-ray, optical, and radio. We consider a range of possible emission mechanisms for BNS mergers; some of the possible pathways and end products are illustrated in Figure~\ref{fig_mm_scenarios}. Furthermore, our investigation uses the full compliment of GW detector network configurations that will be available through the advanced detector era by including KAGRA, LIGO-India and a proposed southern Hemisphere detector AIGO.

The paper is organized as follows: in section ~\ref{method} we describe the framework we use to simulate the GW detection and localization and provide these results in section~\ref{result}.
In section~\ref{telescope} we discuss the potential of EM telescopes for rapid follow up observations and in section \ref{sec_astro} provide a review of some of the possible EM emissions that could be captured by prompt observations. Then in section \ref{sec_astro_flux} we use our simulated detection and localisation statistics with observational data to determine the astrophysical implications of low-latency GW detection and prompt EM follow up. We then summarise our findings in section~\ref{concl}.
\begin{table*}
   \caption{The workflow of a low-latency CBC detection pipeline showing the latencies achieved in science and engineering runs as well as a projected optimistic latency. The LIGO S6-Virgo VSR2/VSR3 joint observing run was performed during 2009-2010;  ER5 and ER6 were engineering runs used to streamline data analysis pipelines by simulating the environment of a real science run. $\ddag$ This assumes that vetting can be eliminated through automatic processing.}
   \centering
   \begin{tabular}{l|c|c|c} 
   \hline
      Process   & Latency in S6-VSR2/VSR3 joint run & ER5/ER6
      $\dagger$ latency & Optimistic latency \\
      \hline
      \hline
      GW data acquisition, calibration \& distribution & 1 min & $\sim$10s & O(1)s\\
      GW trigger generation & 2$\sim$7 mins & $\sim$40s & O(1)s \\
      Follow-up preparison & 2$\sim$3 mins & $\sim$ 60s & O(1)s \\
      Human vetting & 20$\sim$30 mins & N/A & 0s $\ddag$ \\
 \hline
   \end{tabular}
   \label{latency}
\end{table*}
\section{Methodology}
\label{method}
\subsection{GW Detection}
\label{detection_method}
The response of a laser inteferometric GW detector to a GW signal at a given time $t$ is given by:
\begin{equation}
  h(t) = F_+(t)h_+(t) + F_{\times}(t)h_{\times}(t) \label{eq:strain}
\end{equation}
where $F_+$ and $F_{\times}$ are the antenna pattern functions of two independent plus and cross GW polarizations denoted as $h_+$ and $h_{\times}$.
We adopt the convention of the antenna responses given in \citet{krolak98}.
The antenna patterns are expressed as functions of the source sky location---the right ascension $\alpha$, the declination $\delta$ and the polarization angle $\psi$.
In our simulation, we calculate the waveform $h_+(t), h_{\times}(t)$ using the second order post newtonian (PN) expansion. It is a function of eleven parameters: the luminosity distance of the source $r$, the orbital inclination angle $\iota$, the coalescence phase $\phi_c$, masses and spins (three parameters for each spin) of these two inspiral bodies. In this paper we only consider non-spinning BNS coalescence.

\begin{figure*}
        \centering
         \subfloat[]{
                \label{sky}
                \includegraphics[width=0.37\textwidth]{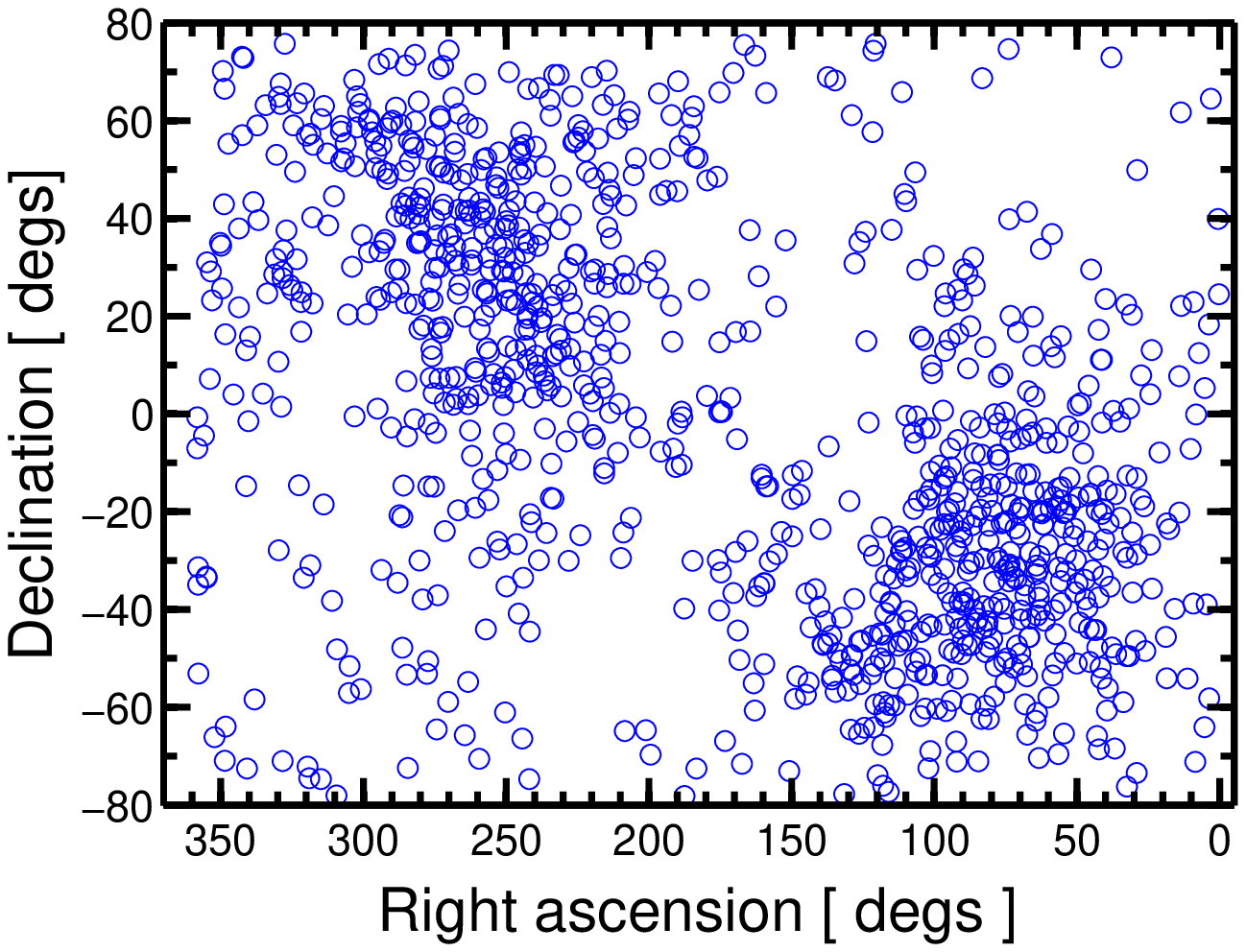}}
          \subfloat[]{
              \label{incl}
                \includegraphics[width=0.35\textwidth]{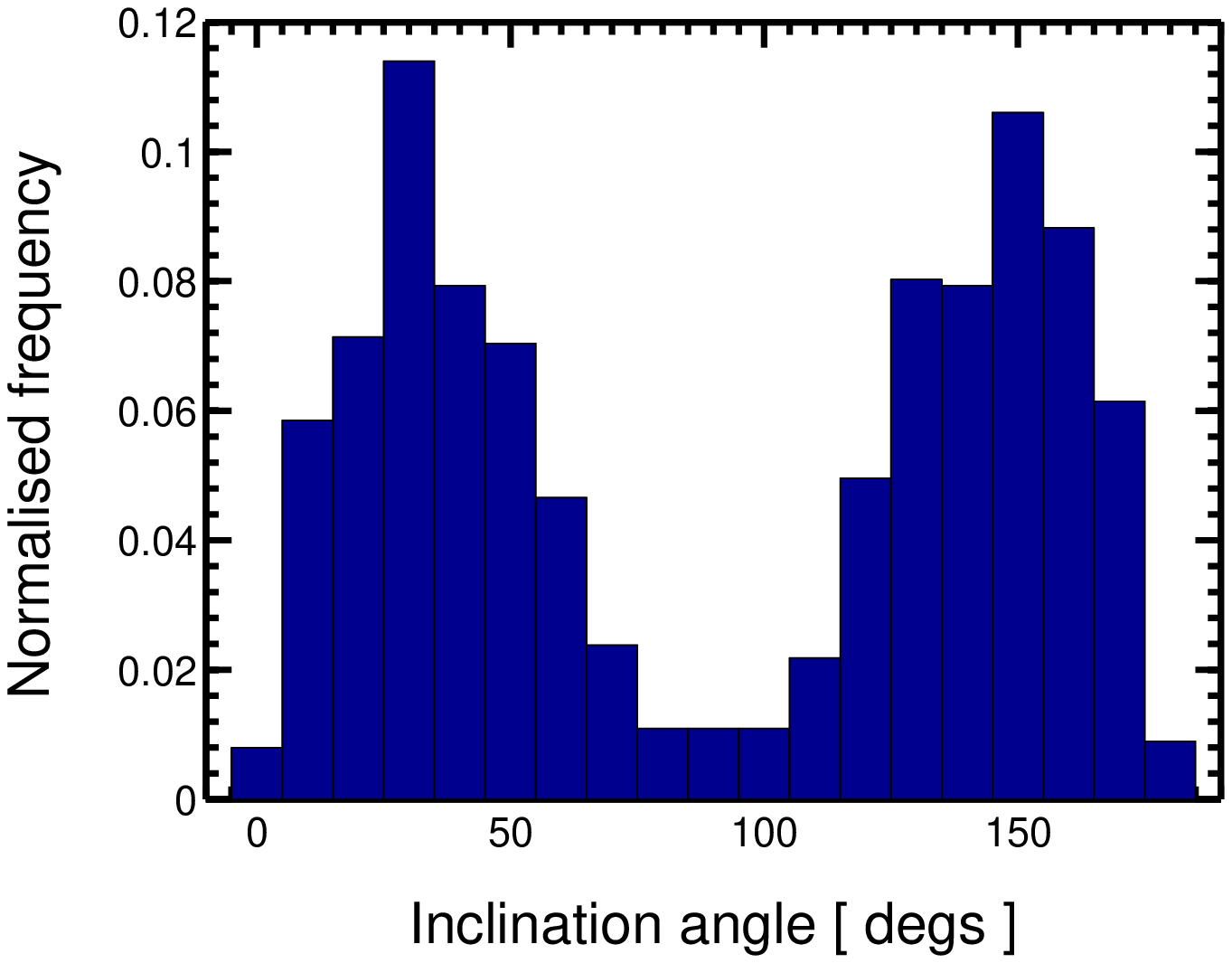}} \\
          \centering
         \subfloat[]{
                \label{chirpm}
                \includegraphics[width=0.35\textwidth]{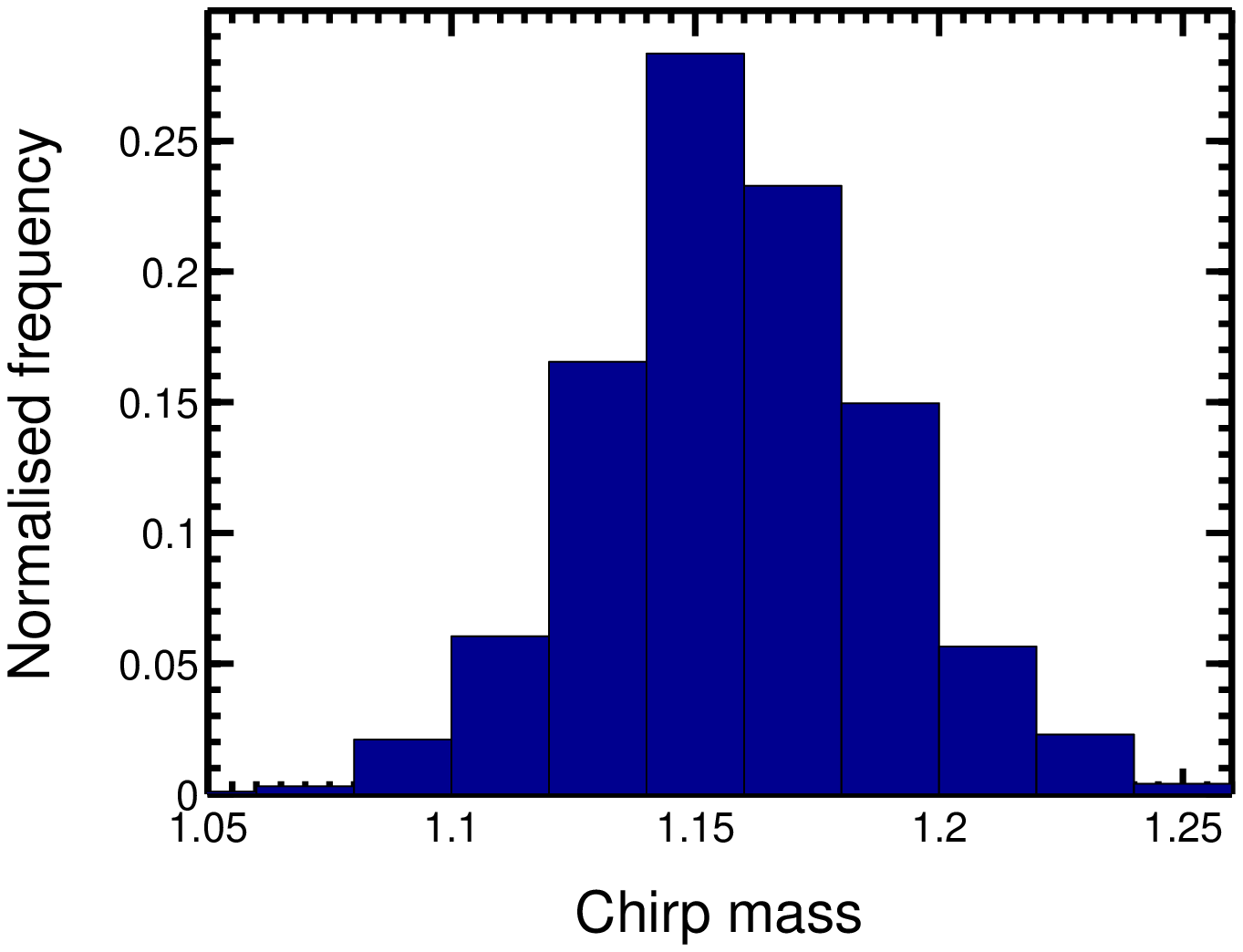}}
          \subfloat[]{
                \label{totalm}
                \includegraphics[width=0.35\textwidth]{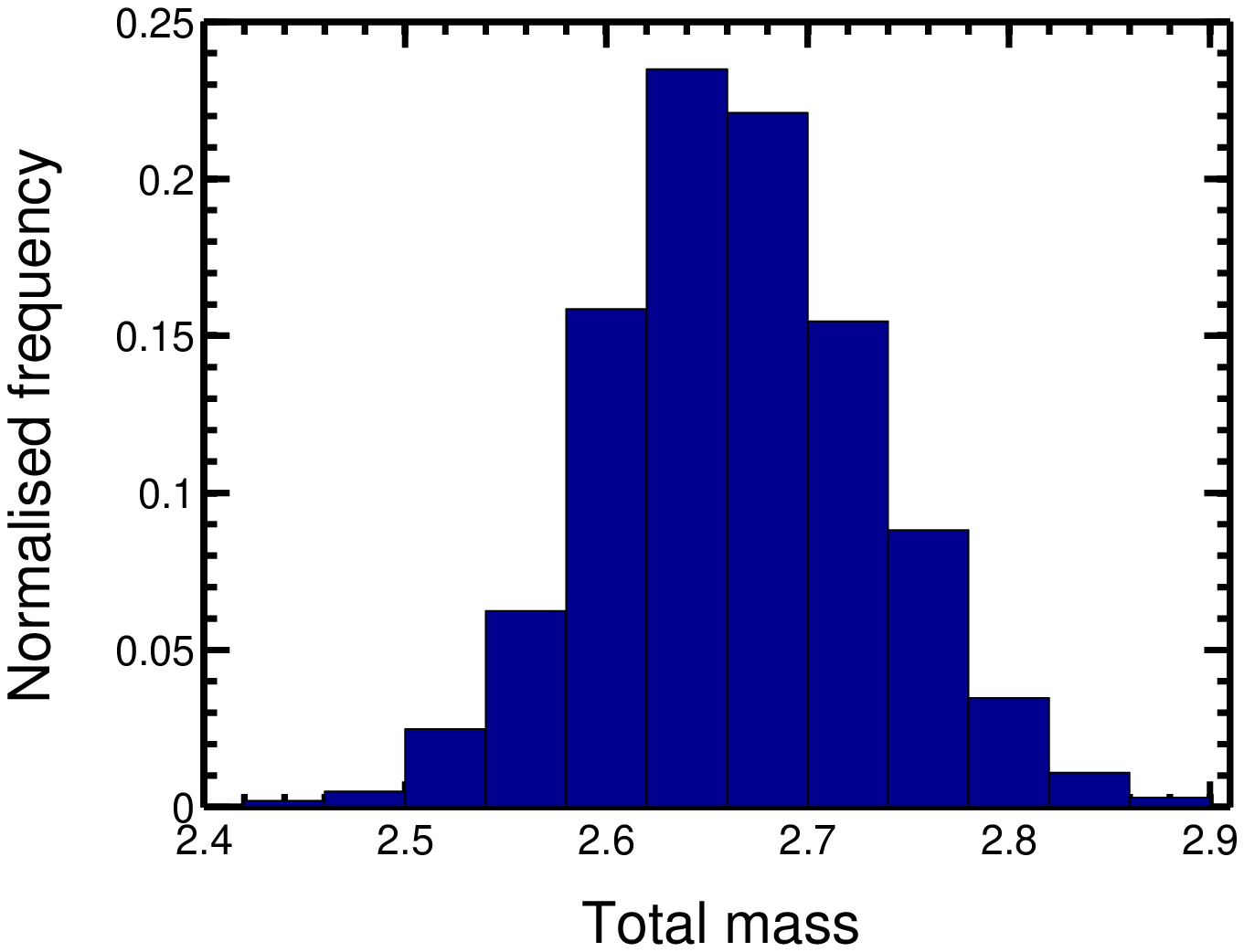}}
                \label{distribution}
        \caption{The parameter distributions for the detected sample of sources. We show the distributions of: (a) sky location;(b) inclination;(c) chirp mass and (d) total mass.}
        \label{distribution2}
\end{figure*}

Matched templates are constructed from the CBC parameter space to perform filtering on the detector data to search for and extract GW signals~\citep{finn92}. The output of matched filtering is the signal-to-noise ratio (SNR), from which a selection criteria is applied to yield credible GW triggers. Throughout this paper we assume that our detections yield an optimal SNR for each source; that is we assume the template matches the source exactly.

For a network of detectors, two strategies can be considered to select GW candidate triggers. The coincident selection criteria (which we refer to as the triggering criteria in the later part of the paper) requires at least two detectors to detect signals above a predefined SNR threshold. A coherent selection criteria assumes that data from different detectors form coherent responses from the source; thus a single statistic is produced and applied against a threshold for trigger generation.

\subsection{Latency of GW Event Triggers}
\label{low_latency_pipeline}
The latency of a GW detection pipeline is defined as the time elapsed from when the source has sufficient SNR for detection to the time it is registered in the GW event database\footnote{\texttt{GraceDB}: The Gravitational-wave Candidate Event Database}. Table \ref{latency} outlines the workflow of a typical low-latency CBC pipeline showing the latencies achieved in the LIGO S6-Virgo VSR2/VSR3 joint observing run, the ER5 and ER6 engineering runs as well as a projected optimistic pipeline.

The workflow was first practiced in the latest joint LIGO/Virgo science run~\citep{abadie11} and the latency is shown in the second column of Table~\ref{latency}. In this process detector data was acquired, conditioned, and distributed. Candidate triggers including CBC triggers were generated by different low latency GW search pipelines and submitted to the database. Significant events were then selected and validated manually to form alerts to participating EM observatories. The total time lag was of order 30 mins of which 20 mins consisted of human validation. The third column of Table~\ref{latency} shows the results of a recent engineering run with low-latency pipelines specific to search for CBC signals and the final column shows estimated optimistic latencies possible in the future.

GW detector data are first acquired onsite and calibrated and distributed to the computing node where search pipelines are running. From the latest engineering runs, this process takes of order 10 seconds. In principle, there is no intrinsic obstacle preventing this procedure to achieve minimal latency. Therefore, we assume in the most optimistic scenario that data can be packed, calibrated and distributed within O(1) s.

At the time of writing, three low-latency GW search pipelines have been implemented for the LIGO-Virgo collaboration to search for CBC signals~\citep{first_low_latency_inspiral, mbta,  cannon12, jing,spiir}. These three search pipelines have all achieved an average of around 40s latency from the time of data acquisition to trigger registration. This latency is mainly due to the processes of data whitening, filtering, and coincident analysis. By employing high-performance computing hardware such as graphic processing units (GPUs)~\citep{shinkee10,liuyuan} and new high-performance algorithms, this latency could be further reduced. We assume in the future O(1)s latency for filtering and trigger generation from multiple detectors can be achieved in the most optimistic case.

The third row of Table~\ref{latency} shows the latencies introduced through follow-up preparation, which includes the gathering of event parameters, e.g. sky localization and calculating telescope tilings. The main latency of this part comes from event localisation. A low latency localization pipeline \texttt{BAYSTAR}\footnote{BAYESian TriAngulation and Rapid localisation} has been developed and tested ~\citep{leo_bayestar} and has achieved latencies in localization of seconds; the run time of this method is shown to be almost perfectly proportional to the number of CPU cores. Thus this latency could be largely reduced and in the optimistic scenario we assume O(1)s latency for this procedure.

Human vetting introduces the highest latency as the first detections needs to be validated carefully. It is natural that once enough detections have been recorded, the knowledge gained will allow this part of the routine to be eliminated by vetoing candidates automatically. We therefore estimate O(1)\,s for the optimistic case.

In summary, we consider several latency scenarios that include both GW detection and localization for GW triggering: an ideal zero latency and latencies of [1s,10s,40s] which takes into account the range of extreme to plausible latencies to send out a GW trigger. In section~\ref{sec_astro}, our discussion is focused on a latency of 40s which is achievable in the coming science runs and an optimistic latency of 1s.

\subsection{Sky localisation accuracy}
To determine the localisation accuracy of the detected GW events, we utilize the formalism of \citet{wen10} who used the Fisher information matrix to derive a geometrical expression for sky localization accuracy. According to the Cram$\acute{e}$r-Rao bound \citep{cramer}, the lower bound of the variance of any unbiased estimator is determined by the inverse of the Fisher matrix $\mathbf{\Gamma}$. Using this property, one can calculate the lower bound on the localisation error area. This estimation is applicable to any unbiased estimators and to all GW waveforms. This method is also used in ~\citet{schutz11} for a sky localization comparison of different detector networks.

\citet{wen10} presented angular resolution formulae for two different detection scenarios: a best case scenario assumed that the waveform was known but the arrival time was unknown; a worst case scenario assumed that both these quantities were unknown. The calculations include both phase and amplitude information of the signal. The angular resolutions derived from the best case scenarios were shown to be around half the areas of those determined assuming the worst case scenario \citep{chuqi12}. In this paper, we assume the best case scenario to calculate angular resolution for each detection. This is given as:
\begin{equation}
\Delta\Omega
  =4\sqrt{2}\pi c^2 \sqrt{\frac{(\sum_i \xi_{i})^2}{\sum_{J,K,L,M}\xi_{J}\xi_{K}\xi_{L}\xi_{M}|(\mathbf{r}_{KJ}\times\mathbf{r}_{ML})\cdot\mathbf{n}|^2}}~,
\label{eq2}
\end{equation}
with $\mathbf{n}$ the direction of the source which is a function of right ascension $\alpha$ and declination $\delta$ and $\mathbf{r}_{KJ}$ the distance vector from detector K to detector J; the quantity $\xi_i=\int _{-\infty}^{\infty}f^2|z|^2 df$ can be interpreted as noise-weighted energy flux received by the $i$th detector and $z$ is the SNR (as discussed in section~\ref{detection_method}).

\subsection{Monte-Carlo Simulation}
\label{Simulation}

To produce a sample of detection statistics we simulate a population of 20,000 BNS sources. Our Monte-carlo code extends other studies by considering a full complement of parameters for the BNS waveform and the sky location.
The values of right ascension $\alpha$ and polarization angle $\psi$ are draw from a uniform distribution within the range $[0, 2\pi]$. Values for the declination $\delta$ angle and inclination angle $\iota$ are drawn from a distribution of cosine values uniformly distributed over the range $[-1,1]$. For each event a source distance $r$ is chosen from within $[10,500]$Mpc; this range encompasses the horizon distances of 428Mpc for aLIGO, 332Mpc for AdV, and 322Mpc for KAGRA. For the component masses of each event we assume a Gaussian distribution based on the study of ~\citet{ozel2012mass}; this study suggests that the mass distribution of BNS systems peaks around $1.33M_\odot$ with a standard deviation of $0.05 M_\odot$.

It is expected that the aLIGO-Livingston (denoted L), aLIGO-Hanford (H), and AdV (V) detectors will reach their full sensitivities around 2019. To model the sensitivities of the advanced LHV detector network, we use the `zero tuning, high power' spectral density noise curve of \citet{SF_aLIGO} for aLIGO-Livingston and aLIGO-Hanford; for AdV we use the spectral density curve of ~\citet{SFaVirgo}.

Our triggering selection criteria requires a SNR of 7.5 in at least 2 detectors \footnote{We note here that the detection pipeline
\texttt{gstlal} is configured to use a single detector threshold SNR of 4.}. This is equivalent to 91\% of the reach of a single aLIGO detector triggering off an SNR of 8; this also translates to a 194 Mpc horizon distance. Assuming there will be tens of detections of an LHV network per year, we present our detection estimates as percentages - this allows the statistics for different thresholds to be easily obtained through scaling.

We obtained a total of 1010 LHV coincident detections from our simulated population of 20,000 sources. For an enlarged detector network we assume the same coincident threshold. The key statistics will be presented in section~\ref{detection_network_larger}. We note that some studies have used a coherent analysis threshold to select sources, e.g.~\citet{schutz11}. The coincident selection thresholds shown here can be simply converted to corresponding coherent analysis thresholds via the detection volume. The statistics of detections from coherent analysis should be similar to the statistics of detections shown in the following section.

\section{Results for GW Trigger Events}
\label{result}
\subsection{Parameter Distributions of Detected GW Events from LHV network}
Fig.2 shows the distributions for the different parameters of the 1010 simulated BNS detections for an LHV network in the advanced era. Fig.\ref{sky} shows that the distribution of detected sky locations is bimodal. The locations of maxima and minima are consistent with the landscape of the antenna power pattern of a LHV detector network \citep[e.g. Fig.2 in][]{nissanke09}. As shown in several studies~\citep{manzotti12,schutz11,nissanke09} adding more detectors to GW search will significantly improve sky coverage.

The distribution of detected inclinations shown in Fig.\ref{incl} is bimodal and shows a detection preference towards low inclination angles. We note however that the fraction of detected inclinations below $25^{\circ}$ is less than one quarter \citep[see also][]{schutz11}. The distributions of the chirp mass and total mass of sources are shown in Fig.~\ref{chirpm} and Fig.~\ref{totalm}.
\subsection{The localisation of BNS detections}

In this section we present the localization evolution statistics for the LHV detections and for an extended detector network that covers the entire parameter space of possible BNS sources. There is a crucial interplay between sending out a prompt alert to an EM facility and surveying the GW error region before the EM source fades. An early trigger can allow a telescope to get on source, but at the expense of a larger GW error region. The situation will vary significantly for different types of telescope and is also dependent of the type of EM emission. In preparation for the later analysis of EM follow-up in Section 6 we have calculated the percentage of sources detectible within different sized error regions at different times before merger. We note that previous studies have calculated the evolution of the localization property with time using a qualitative analysis \citep[e.g.][]{cannon12} rather than the quantitative approach adopted in this paper.

\begin{figure}
        \centering
        \subfloat[]{
                \centering
                \label{snr}
                \includegraphics[width=0.45\textwidth]{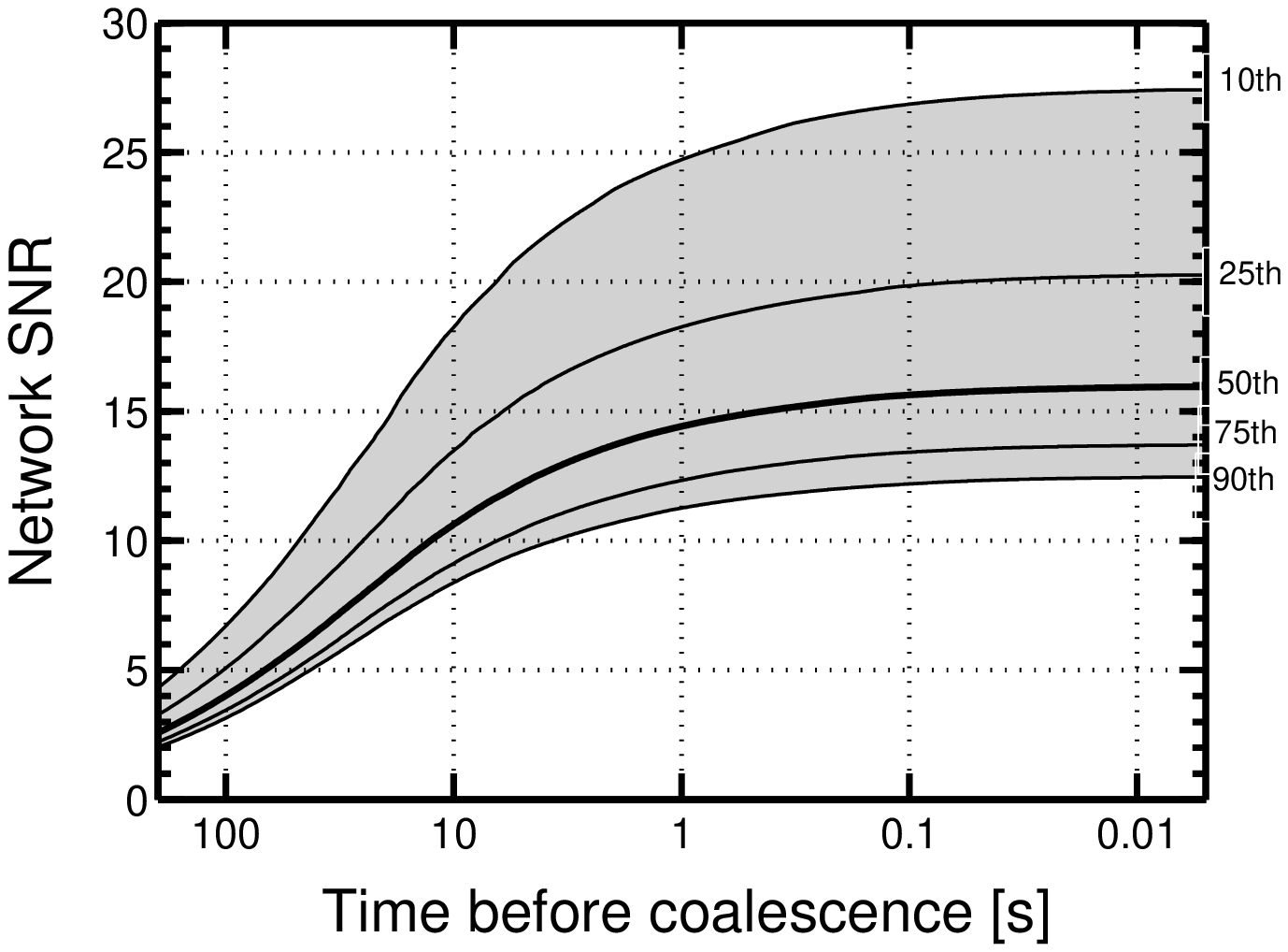}}\\
                \quad
        \subfloat[]{
                \centering
                \label{ar}
                \includegraphics[width=0.45\textwidth]{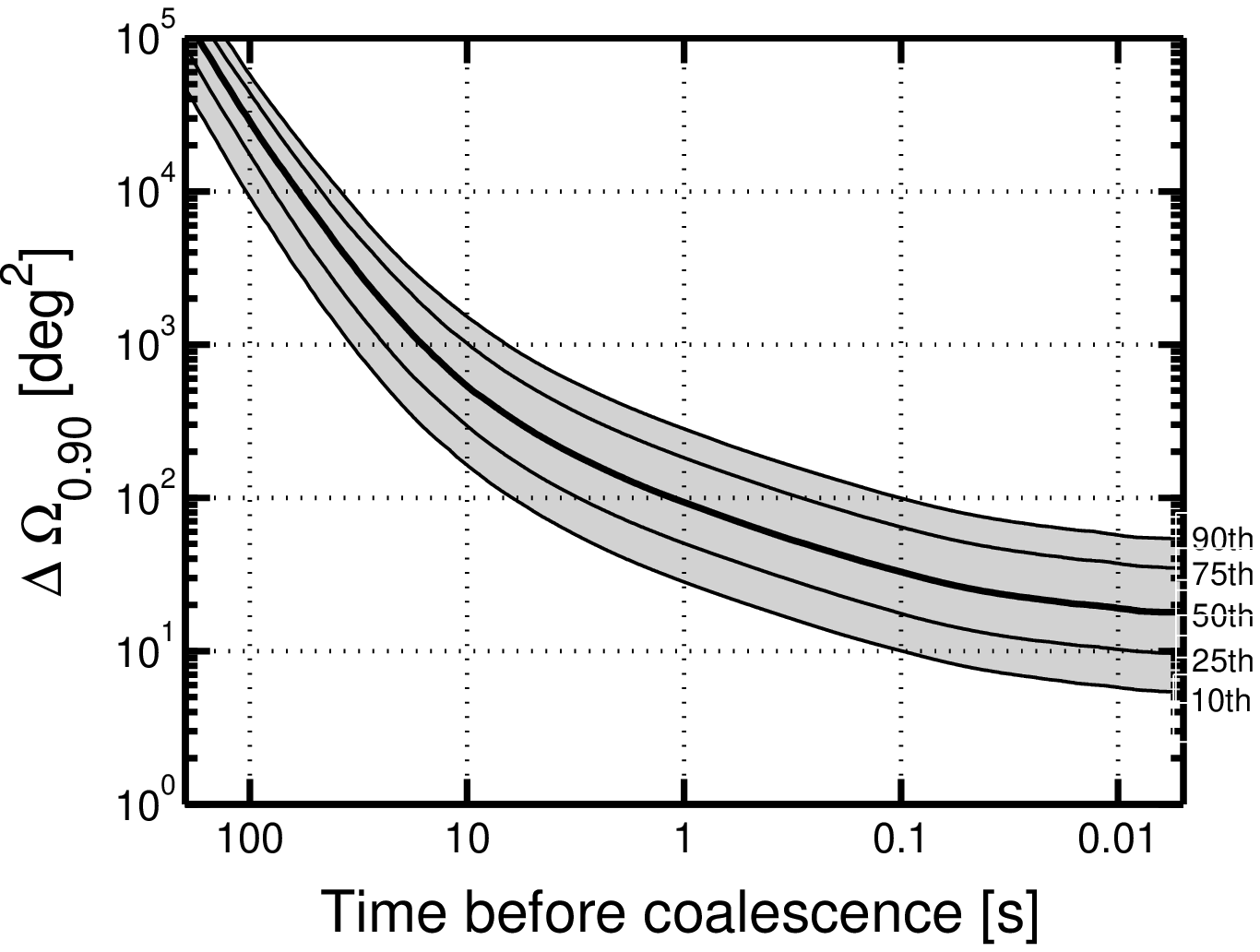}}
        \caption{Percentile curves of (a) network SNR and (b) $\Delta \Omega$ (error box at 90\% confidential level) as a function of time before merger for the simulated BNS triggers obtained by an LHV network.}
        \label{ew}
\end{figure}

Figure \ref{ew} displays the percentile charts for the evolution of network SNR (\ref{snr}) and localisation error area (\ref{ar}) for the 1010 LHV detections as they approach merger (t=0). The network SNR is calculated as root-sum-square of individual SNRs. The plot shows that the median network SNR (thick solid line) is less than 5 at 100s before coalescence. This SNR increases by 1 every 10 seconds during -100s to -10s and by 0.5 every second during -10s to -1s before reaching a SNR of 16 at merger. We find that 10\% of the detections will yield a SNR greater than 27 at merger time which corresponds to a localisation area of around 5 deg$^2$.

Figure~\ref{ar} shows the percentile range of localisation error areas at 90\% confidence level as a function of time before merger. On average, the angular resolution is about $30000~\mathrm{deg}^2$ at -100s. As the network SNR increases, this improves to $< 5000~\mathrm{deg}^2$ at -40s, $<\,500~\mathrm{deg}^2$ before -10s and less than a hundred square degrees at -1s. These results support the early localisation results of a number of recent studies: see for example \citet{cannon12}, \citet{manzotti12} and Ch.3, Fig 3.3 of  \citet{Singer_2015PhDT}.

\subsection{Improved Early Warning Events with a Larger Detector Network}
\label{detection_network_larger}
\begin{figure}
  \centering
   \includegraphics[width=0.47\textwidth]{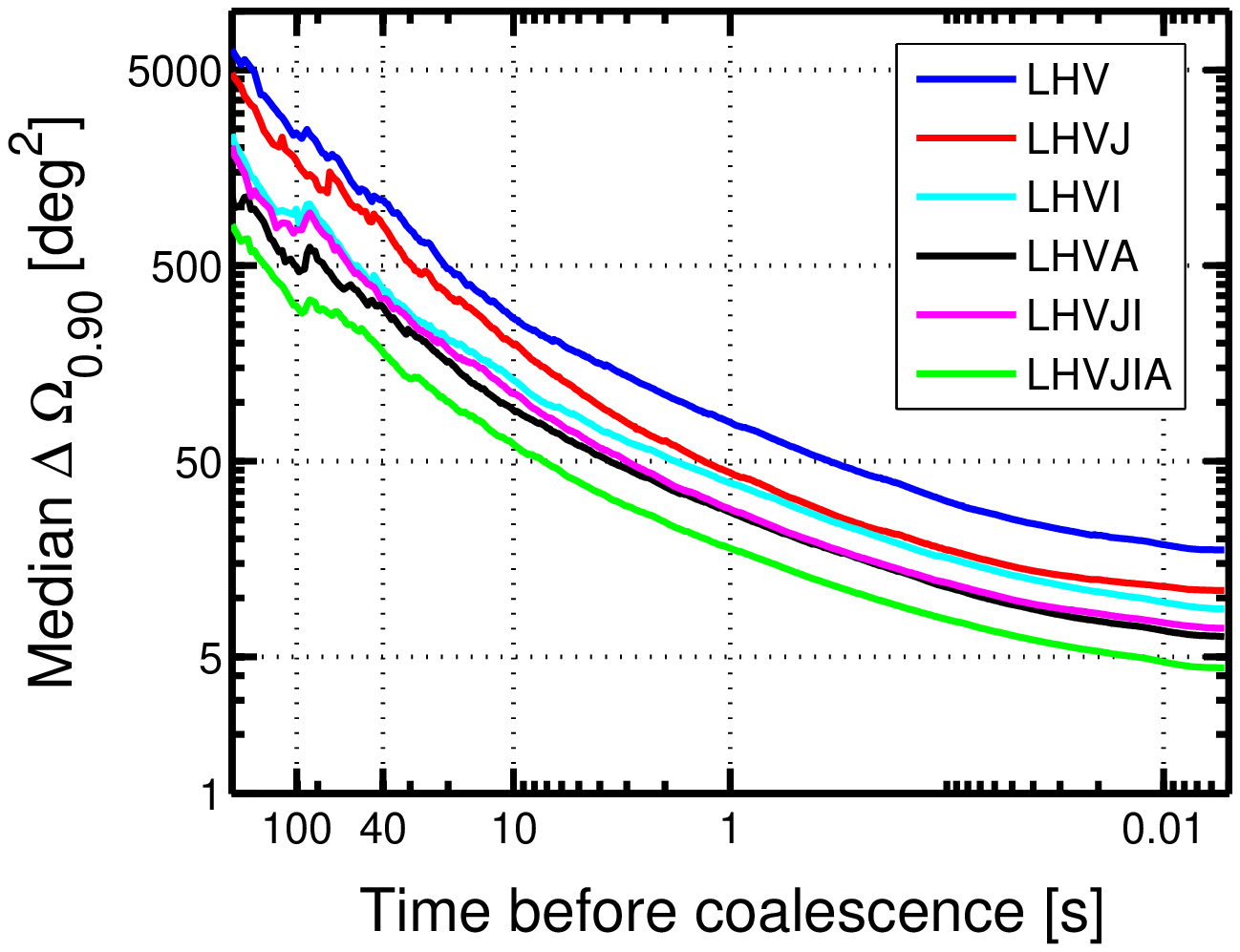} 
   \caption{Median localisation error areas of detections at exact times prior to merger for different detector networks. The fluctuations seen at early times are due to the smaller number of detections in that regime.}
   \label{trig}
\end{figure}

In this section we investigate the benefits of an enlarged GW network in terms of early detection and localisation; the scenarios considered are applicable to the advanced detector era beyond 2019. We consider a number of additions to the LHV network: these are KAGRA (labeled as J), the proposed Indian detector: LIGO-India (labeled as I), and AIGO (labeled as A). Though AIGO is not approved financially at the moment, it is considered here for comparison. As in section \ref{Simulation} we use the aLIGO spectral density for LIGO-India and AIGO, and the spectral density from~\citet{kagra} for KAGRA. The main results are given in Table \ref{early_loc} and Figure~\ref{trig}.
\begin{table*}
   \caption{The detection statistics obtained for different advanced detector networks shown in the first column. The second column shows the relative detection capability of each network in comparison to LHV. Columns 4 to 7 provide two sets of statistics at different times before merger for each detector network: in the first row we provide the percentage of detections achieved pre-merger; the second row gives the sizes of the error regions required to contain 10\%, 50\% and 90\% of the detections at exact times before merger. $\ddag$ This number is lower than that of LHVI/LHVA network as the sensitivity we adopted for detector J is 0.75 of that of detector I/A.}
   \centering
   \begin{tabular}{l|p{2cm}|p{5cm}|c|c|c|c} 
   \hline
   \hline
    Network   & Detection capability & & \multicolumn{4}{c}{Time before merger} \\
            & & & 40s & 10s & 1s & 0s \\ \hline
      LHV  & 1.00 & Percentage detected & 9\% & 35\% & 76\% & 100\% \\
            & & [10\%/50\%/90\%] error regions [deg$^2$] & 300/1000/2200 & 70/270/780 & 20/80/240 & 5/20/50 \\ \hline
      LHVJ & 1.12 $\ddag$ & Percentage detected  & 8\% & 32\% & 76\% & 100\% \\
            & & [10\%/50\%/90\%] error regions [deg$^2$]  & 240/820/1600 & 50/200/470  & 10/40/120 & 3/10/30\\ \hline
      LHVI  & 1.28 & Percentage detected  & 9\% & 33\% & 75\% & 100\% \\
            & & [10\%/50\%/90\%] error regions [deg$^2$]  & 120/380/910 &    40/130/350 & 10/40/100 & 3/9/20\\ \hline
      LHVA  & 1.27 & Percentage detected  & 9\% & 35\% & 76\% & 100\% \\
            & & [10\%/50\%/90\%] error regions [deg$^2$]  & 80/310/850  & 30/90/250 &      9/30/80  & 2/6/20\\ \hline
      LHVJI & 1.35 & Percentage detected  & 8\% & 32\% & 75\% & 100\% \\
            & & [10\%/50\%/90\%] error regions [deg$^2$]  & 100/340/830 & 30/110/300    &  9/30/80  & 2/7/20\\ \hline
      LHVJIA  & 1.57 & Percentage detected  & 9\% & 33\% & 75\% & 100\% \\
            & & [10\%/50\%/90\%] error regions [deg$^2$] & 60/180/400 & 20/60/140 & 6/20/40 & 1/4/10 \\ \hline
   \end{tabular}
   \label{early_loc}
\end{table*}

The second column of Table~\ref{early_loc} shows the detection capability of different detector networks in terms of the number of detections normalized to that obtained from a LHV network (the reference detection capability of LHV is given here as 1.00). For larger network configurations, we find that the detection capability is improved considerably. Using the same coincident triggering threshold for different detector networks, we obtain 1133 detections for LHVJ, 1291 detections for LHVI, 1366 detections for LHVJI, 1280 detections for LHVA, and 1584 detections for LHVJIA.  The detection capability of a larger detector network depends on the sensitivities of the additional detectors. We find that adding a single detector with the same detection capability as aLIGO to the LHV network will increase the detection rate by about 30\% while adding KAGRA to the LHV network will improve the detection rate by 12\% as the sensitivity of KAGRA is 0.75 of that of aLIGO. The detection percentage for each network at different times of the final merger phase is shown in the first row for each network configuration.

Table \ref{early_loc} also shows localisation achievable at exact times before merger for different GW network configurations. We illustrate this by showing the sizes of the error regions required to contain 10\%, 50\% and 90\% of the detections. All the error region numbers are estimated at the $90\%$ confidential level. We find that adding a KAGRA or LIGO-India instrument to the LHV network can improve the localisation by a factor of 2. Increasing the network to five detectors can lead to a factor of 3 improvement. Adding AIGO on its own results in an improvement that is equivalent to adding both KAGRA and LIGO-India due to the longer baseline and optimal location. Adding AIGO to the 5-detector network, can further improve the localisation by a factor of around 2 -- this is a factor 5 improvement over the initial LHV network.

An important fact is that only a fraction of detections can be detected at times prior to merger. We take this into consideration and show the localisation error area for the detections at different times prior to merger. Fluctuations of localisation error area at early times in Fig.~\ref{trig} are due to small number of detections in that epoch. We  note our estimated localisation error areas determined near merger time are comparable, though slightly more optimistic than~\citet{fairhurst11} and \citet{ligo_localization}. Whilst the aforementioned studies used the Fisher matrix for the triangulation localisation method, our estimates are based on the Fisher matrix regardless of the localisation method, thus represent a lower limit of all non-biased estimations.

In practical terms, the latencies introduced in the GW-EM effort will affect the timing of the EM followup observation and therefore constrain the phenomena we are able to observe. Using Table~\ref{early_loc}, one can easily consider different GW-EM reaction latency scenarios and find the detection ratio and localisation error area for follow-up observations at merger or at post-merger timescales.
For example, assuming a 40 second total reaction latency -- incorporating the whole GW trigger generation and EM-followup readiness latencies -- can be achieved, those GW events that pass the GW triggering criteria 40s before the merger can potentially be observed right at their merger time by EM instruments with wide FoVs and rapid responses. For the LHV network, which is expected to be operational at early stage of advanced era, around 9\% of all detections can be observed at the merger time within 40s. The median localization error area of the detections at the merger time is over 1000 deg$^2$.  The median localisation error area can be brought down to about 400 deg$^2$ by an LHVI detector network and 10\% detections of LHVI network at the merger time could be localised within 120 deg$^2$.  With a five-detector network, 10\% of the detections could be discovered and localised within 100 deg$^2$ at the time of merger.

If a greater effort could reduce the total reaction latency to 10 seconds (this highly optimistic scenario would require an EM facility to be on source), the detection ratio at the merger time will be raised to around 33\% for the six networks which is three times the observed detections at 40 second latency. More importantly, the localisation accuracy is also three times better than that achieved with 40 second latency. For the LHV network, the median localization area is improved 4-fold to 270 square degrees and 10\% of detections could be localised within 70 deg$^2$. Adding KAGRA or LIGO-India into the LHV network will improve the localisation to less than 50 deg$^2$ for 10\% of their detections. The localisation accuracy will be further brought down to about 30 deg$^2$ for a network of LHVA or LHVJI of 10\% of their detections.  The best localisation accuracy comes at an operation of six detectors with around 90\% of the detections localised to within 140 deg$^2$. We consider further the possibilities of existing EM facilities in the next section.

\section{EM Telescopes for Prompt GW Follow-up Observations}
\label{telescope}
In this section we will discuss some of the EM instruments in different EM wavelength bands that will be able to facilitate prompt GW follow-up -- we consider a selection from those that currently have MoUs for aLIGO/AdV follow-ups in place. We focus first on the high energy which is traditionally low latency (usually employing triggering instruments) but will not be fast enough unless there is an agreement for prompt Target of Opportunity (ToO) in place; we note that high-energy observations of GRBs over the last two decades have produced significant science returns in this domain. We then discuss optical telescopes which is proven for low latency through the successful use of fast response instruments for GRB follow-up. Finally we consider radio which though electronic steering and/or rapid slew-times, can make rapid response follow-ups achievable.
\begin{table*}
  \caption{The properties of some of the high-energy instruments planned to be operational during aLIGO.
  [1] \citet{Gehrels_Swift_2004}; [2] \citet{Meegan_Fermi_2009}; [3] \citet{Bartos2014MNRAS}; [4]  \citet{MXRT_2014}.  NOTES: $\S$ Sensitivity approximated using \citet{Zhang2013ApJ} equation (3) $F_{\mathrm{x,th}}=6 \times 10^{-12}/T_{\mathrm{obs}}$ \uflux and assuming $T_{\mathrm{obs}}$ = 10s. $\dag$ Sensitivity over 100s estimated using Fig 2. of \citet{Atwood2009ApJ} and converted to
   \uflux \, using a standard band function with $\alpha=-1$, $\beta=-2.25$ and $E_{\mathrm{P}}=511$ keV. $\flat$ Approximated using \citet{Funk2013APh}. $\natural$ Sensitivity in survey mode based on [3] in which exposure time includes an estimate  of the required slewing times to tile a 1000 deg$^{2}$ area using convergent pointing mode.
   }
\begin{center}
  \begin{tabular}{ccccccc}
\hline
\hline
  Instrument &      Energy Range   &FoV          &  Sensitivity  & Exposure Time  & Response Time &  Ref\\
            &                     & [deg$^{2}$] & [\uflux]       &               &               &     \\
\hline
\hline
\emph{Swift} - BAT  &  15-300 keV  & 4600     & $1.2 \times 10^{-8}$ & - &  Coincident Observation       & [1]\\
\emph{Swift} - XRT $\S$  &  0.3-10 keV & 0.15  & $6 \times 10^{-12}$  &  10  &  1-2 hours  &    [1]\\
Fermi - GBM       &  8\,keV --40\,MeV                    & 30000       & $4 \times 10^{-8}$ &  -  & Coincident Observation   &    [2]\\
Fermi - LAT$\dag$       &  0.02-300\,GeV                     &  8000      & $1.4 \times 10^{-7}$ & 100s   & 30s post GBM trigger       & [2]\\
CTA $\flat$              & 0.03 -- 100 TeV  &6 -- 8\,deg$^{2}$     &  $6 \times 10^{-9}$@ 25GeV   & 1000s  &   20--60secs         & [3] \\
CTA (Survey mode) $\natural$& 0.03 -- 100 TeV   &$\sim$1000\,deg$^{2}$  &   $6 \times 10^{-8}$@ 25GeV  & 1000s  &   20--60secs         & [3]\\

H.E.S.S. $\flat$ &  0.05 -- 20 TeV     & 15    & $\sim 6 \times 10^{-8}$ @ 25GeV  & 1000s & 30s       & [3]\\
SVOM - ECLAIRS &  4-250 keV      &  89       &  $7.2 \times 10^{-10}$    &    1000s  &        $\sim$ 1-2 hours       & [4]\\
SVOM -  MXT     &  0.2-10 keV     &  64 arcmin &  $5.6 \times 10^{-11}$    & 10s  &        $\sim$ 1-2 hours         & [4]\\
\hline
\end{tabular}
  \label{table_high_energy}
  \end{center}
\end{table*}

\begin{table*}
  \caption{The properties of a selection of the optical telescopes planned to be operational during aLIGO
   [1] \citet{Klotz_2013}; [2] \citet{Keller2007PASA};[3] \citet{Coward2010PASA}; [4] \citet{Hodapp2004}; [5] \citet{Comm_GOTO_2015}; [6] \citet{Ghosh2015}; [7] \citet{Smith_2014SPIE,Bellm_2014} [8] \citet{Paul2011CRPhy} }
\begin{center}
  \begin{tabular}{ccccc}
\hline
\hline
Telescope              &FoV          &  Limiting Magnitude    & Exposure                                      & Ref\\
                       & [deg$^{2}$]&      R-band            &                                              &   \\
\hline
 TAROT                &    3.5          &        18               &  60                                          & [1]\\
 SkyMapper            &    5.7          &        21               &  110s                                          & [2]\\
Zadko                  &    0.15        &        21               &  180s                                         & [3]\\
Pan-STARRS               &   7.0       &        24               &  30s                                         & [4]\\
GOTO               &   18.0-36.0       &        21               & 5m                                        & [5]\\
BlackGEM               &   40.0       &          22             &  5m                                         & [6]\\
Zwicky Transient Facility (ZTF)       &   47       &              20.5-21          &  30s                                        & [7]\\
Ground based Wide-Angle Camera (GWAC)       &   8000       &            16         &  10s                                      & [8]\\
\hline
\end{tabular}
  \label{table_optical}
  \end{center}
\end{table*}

\begin{table*} 
\caption{The capacity of radio telescopes in consideration [1] \citet{Tingay2013PASA}; [2] \citet{Murphy2013PASAa}; [3] \citet{vanHaarlem2013AA} $\dagger$; [4] \citet{Oosterloo2010}; $\natural$ APERTIF will be able to cover 100\,deg$^2$ at lower sensitivity in a ``fly's eye'' mode.}
\begin{tabular}{c|c|c|c|c|c}
\hline
\hline
Name& Frequency Range &FOV (sq-degs) & limiting flux density &Response Time  & Ref \\
    &                 &[sq-degs]     &                            &               &  \\
\hline
MWA&80-300 MHz&610\,deg$^{2}$@150\,MHz& 20\,mJy & $<$ 10s of secs   & [1]\\
ASKAP&700\,MHz-1.8\,GHz&30\,deg$^{2}$@1.4\,GHz& 640$\mu$Jy beam$^{−1}$ (10s int)&$<$ mins  & [2] \\
LOFAR LBA (Inner)&10-90 MHz&450\,deg$^{2}$@60\,MHz& 5 mJy @60\,MHz  & $<$ min &  [3]\\
LOFAR HBA (Core)&110-250 MHz&48\,deg$^{2}$@180\,MHz& 0.6 mJy @180\,MHz & $<$ min &  [3]\\
APERTIF&1-1.7 GHz&8\,deg$^{2}$@1\,GHz  ($>$100\,deg$^{2}$)$\natural$ & 0.1 $\mu$Jy & &  [4]\\
\hline
\label{radio_instruments}
\end{tabular}
$\dagger$ LOFAR sensitivities quoted for an observation comprising of a total bandwidth of 7.8 MHz
and a 1 hour integration time assuming 6 \\
sub-array pointings are created, calculated
using \url{http://www.astron.nl/~heald/test/sens2.php}
\end{table*}

\subsection{X-ray and $\gamma$-ray follow-ups}
Table \ref{table_high_energy} shows the various properties of some of the high energy instruments with LIGO/AdV MoUs in place.

Space based $\gamma$-ray follow-ups will be led by Fermi and \emph{Swift}. These instruments provide the capability for the discovery of simultaneous EM counterparts; in this respect the large FoV of the Fermi-GBM, which sees almost the entire unocculted sky can be advantageous.
A typical follow-up procedure will be to search through the data for a coincident signal from a location of the sky consistent with the LIGO error region. Assuming coincident [GW/Fermi onboard] triggers, the response time will be around a few minutes. If there are no on-board triggers, a sensitive search through the Fermi data set will require of order 4-12 hours. The Fermi-LAT can provide more information of the high energy afterglow. Typically the LAT autonomously re-points after a GBM trigger with a processing delay of around 30s. If no on-board trigger is available, a ground based search of the data can be initiated; the latency will depend on the time between the GW candidate and the data downlink\footnote{The communication going from the satellite to the ground} and processing, of order one to a few hours.

The \emph{Swift} $\gamma$-ray instrument is a wide-field (1.4sr) coded mask hard X-ray imager (Burst-Alert-Telescope; BAT). Although the \emph{Swift}-BAT is more sensitive than the Fermi-GBM, it observes softer $\gamma$-rays; this last point is important for Fermi-GBM, follow-ups as SGRBs associated with BNS mergers have harder spectra than LGRBs.

In terms of identifying transients, follow-ups in the X-ray have the advantage of a less crowded sky than that of the optical. For X-ray follow-ups the \emph{Swift}-XRTs rapid slew time is particular advantageous; this has resulted in this instrument accumulating around 200 GRB redshifts though co-ordinated observations with ground based telescopes. The \emph{Swift}-XRT is able to respond rapidly to a BAT detection of new transient, localising to within 1-3 arcminutes and then distributing the position via the Gamma-ray Coordinates Network (GCN) within a 20 second window.

\emph{Swift} demonstrated the feasibility of rapid high energy ToO follow-ups by conducting the first multi-wavelength follow-up observation during the 2009–-2010 LIGO/Virgo science run \citep{Evans2012ApJS}. \emph{Swift} followed-up and observed the sky locations for two candidate GW events within 12\,hrs, using a specially designed strategy to target the most probable positions that the GW event originated from. To improve ToO requests with large positional uncertainties, \emph{Swift} implemented tiled XRT observations in late 2011. However for this instrument, at the present time, a response time to ToO of order 1-2 hours \citep{Kanner2012ApJ} by the on board X-ray telescope (XRT) effectively rules this instrument out in terms of low latency follow-ups. If however, in the future, aLIGO could send triggered alerts directly to \emph{Swift} without human intervention, a slew of order 10s of seconds would result in early follow-up observations of the X-ray light curve.

In addition to the \emph{Swift} XRT, Astrosat\footnote{\url{http://astrosat.iucaa.in/}}, an Indian multi-wavelength satellite is was successfully launched in September 2015. Astrosat has instruments covering the UV (UVIT), Optical, Soft \& Hard X-ray (the Cadmium Zinc Telluride Imager; CZTI). For follow-up operations, once a counterpart is identified, re-pointing will bring the target within the FoV of the other instruments for simultaneous coverage from optical though to hard X-ray. For low-latency follow-ups, CZTI, the hard X-ray coded aperture mask instrument with a 36 deg$^{2}$ FoV will be the most useful instrument. As contact with the spacecraft is designed to take place once in an orbital period of 100 minutes, this is typically is the fastest time scale over which a slew can be
commanded; ToO decision making may also be added to this response time. However, one low-latency scenario exists, in that if there was significant hard X-ray/$\gamma$-ray flux in the (100-1000) keV band, the CZTI would detect the flash without re-pointing. Over this energy range the  CZTI acts like a open detector with a 2$\pi$ FOV; we note however that there will be no localization capability at these energies.

Ground based Gamma-ray detectors can also contribute to the GW follow-up effort through the Cherenkov Telescope Array \citep[CTA;][]{Dubus2013APh}. The CTA is a next generation ground-based instrument that will begin construction around 2016, with full operation expected by 2020. CTA will improve over previous experiments ( H.E.S.S, VERITAS\footnote{\url{http://veritas.sao.arizona.edu/}} and MAGIC\footnote{\url{http://magic.mppmu.mpg.de/}}) through increased sensitivity, angular resolution ($\sim 2^{'}$ at TeV), wider energy coverage ($\sim 30$GeV -- 300TeV) and a larger field of view (6\degree-8\degree). This project will consist of two arrays: a southern hemispheric array focusing on Galactic sources and a northern hemispheric array on extragalactic. \citet{Bartos2014MNRAS} has recently highlighted some of the properties that make this instrument ideal for GW follow-up. These include the capability to observe in a pointed mode with a relatively small FoV and in survey mode covering a larger area large sky area ($\sim 1000$ deg$^{2}$ ); fast repositioning is of the order of a few seconds \citep{Doro2013APh}. For low-latency follow-ups the ability to respond to ToO requests and start monitoring the GW error ellipse within 30s is significant \citep{Dubus2013APh}.

CTA is designed to have three types of telescopes with different mirror sizes to cover the full energy range - large scale telescopes (24m) covering $<$ 300 GeV, medium scale (10-12m) covering 100 GeV-TeV and small scale (4-6m) covering higher energies $>$10 Tev. Typically, extragalactic background light (EBL), which is all the accumulated radiation in the Universe due to star formation processes, will attenuate very high energy photons coming from cosmological distances through pair production with EBL photons. This produces the effect in that only the large scale telescopes will be triggered. However, for the case of a GW source within 200 Mpc, TeV photons are not expected to be annihilated \citep{Bartos2014MNRAS}; thus all 3 types of CTA telescopes CTA could be triggered \citep{Comm_CTA_2015} providing a greater energy coverage.

H.E.S.S.\footnote{\url{http://www.mpi-hd.mpg.de/hfm/HESS/}}, an array of 5 Cherenkov telescopes (with 4$\times$ 12m and one 28m diameter mirrors) located in Namibia, has been operational since 2004. This telescope has a mean time to get on source from a random observation position of about 30 seconds \citep{Lennarz2013}. The capability of H.E.S.S. for GW follow-ups has been proven through prompt observations of GRB triggers. A follow-up observation of GRB 070621 was achieved within 7 mins \citep[][]{Aharonian2009A&A} and GRB 100621A was observed within 10 mins.
\subsection{Optical follow-ups}
Optical follow ups have been hugely successfully in the last decade to discover the afterglows and host galaxies of GRBs. For \emph{Swift} follow-ups, this strategy has relied on the prompt localisations of the XRT and the rapid distribution of coordinates via the GCN network. Recently, robotic optical telescopes have been able to identify the optical afterglows of Fermi GBM triggers that have large positional uncertainties (e.g. GRB 130702A; \citealt[][]{Singer2013, SingerPTF2015ApJ}). A successful low-latency EM follow up to a GRB could potentially capture an optical flash or the earliest stages of a particularly bright rising afterglow.

Table \ref{table_optical} lists the capabilities of a number of optical instruments that will be available during the era of advanced GW detectors. Capturing an EM counterpart with such a fast response will be dependent on a number of factors related to the telescope. The FoV of the instrument will determine how quickly the GW localisation area can be surveyed (i.e. the number of tiles required); any subsequent detection will then be dependent on the limiting magnitude or sensitivity of the instrument and on the required exposure time to obtain the required sensitivity \citep[for an in depth discussion see][]{Coward2014}.

For prompt follow-up, the slew time of the instruments will be highly important; for telescopes with lower sensitivities a rapid slew time can compensate through the ability to achieve pointing during the brightest part of the event. However, we note here that for optical instruments, a fast slew time is usually associated with a small FoV.

Fast response optical telescopes such as TAROT have the ability to slew to a target position within 10 seconds. However, the FoV $\sim$ 3.5 deg$^{2}$ and exposure time of around 60\,s mean that time must be allowed to image the full error region. Assuming a 60 deg$^{2}$ error box, corresponding to that of a LHVJIA network at 50\% confidence, using the framework presented in \citet{Coward2014}, we find the imaging time will be of order 17 minutes. Larger FoV telescopes would be able to minimise the required imaging time. For example, Pan-STARRS, a system of four automated 1.8m optical telescopes based in Hawaii has a FoV of $\sim 7$deg$^{2}$ and an exposure time of 30s yielding a 4 minute response for the same error box. The Zwicky Transient Facility (ZTF\footnote{\url{http://www.ptf.caltech.edu/ztf}}) will be an extension of the Palomar Transient Factory (PTF; 7 deg$^{2}$ FoV) with an increased 47 deg$^{2}$ FoV and an order of magnitude improvement in survey speed. GOTO\footnote{\url{http://goto-observatory.org/}}, a dedicated optical follow-up instrument with a proposed initial configuration of 18 deg$^{2}$ instantaneous FoV will require an exposure time of around 5 mins to reach 21 mag \citep{Comm_GOTO_2015}; however, for this instrument a shallower inventory of the large error ellipses may be possible. Additionally, BlackGEM\footnote{\url{https://astro.ru.nl/blackgem/}}, expected on-line from 2016, aims to cover 40 deg$^{2}$ in a seeing-limited configuration with an exposure time of 5 mins \citep{Ghosh2015}.

Considering the FoVs given in table \ref{table_optical}, it is immediately apparent that for fast response observations, unless a wide FoV instrument is on-source, one may require a larger network than LHV to decrease the localisation error \citep{manzotti12} and thus, reduce the number of tiles required to image the full GW error box.

In addition to the facilities outlined in Table \ref{table_optical}, a number of other facilities are in development for GW follow-up and could be operational during the later era of aLIGO/AdV. One such facility is RAMSES which will consists of a set of 16$\times$40cm telescopes and will be equipped with rapid cameras capable of readout times from 100ms to several minutes; it will be able to reach magnitudes of 15 in 1s and mag 20 within around 10 mins. Each telescope has a 2.5 deg$^{2}$ FoV, resulting in a combined 100 deg$^{2}$ FoV; this in combination with a slew time of around 5s would enable instantaneous coverage of a large proportion of a typical aLIGO/AdV error region. RAMSES has received funding and will be based in National Aures Observatory in Algeria\footnote{\url{http://www.craag.dz/}} with a prototype built in France.

\subsection{Radio follow-up}
Table \ref{radio_instruments} provides specifics for some of the radio instruments that have signed MoUs with LIGO/AdV. A period of significant investment in facilities and infrastructure and advances in high-speed computing over the last few years will enable this band to make a telling contribution to time-domain astronomy. The instruments included in this table have the potential to be leading transient detectors by virtue of their large FoV, high sensitivity and significantly for low-latency follow-ups, the capability to respond from 10s of seconds up to within a minute. The triggering methodology has already been implemented by the Arcminute Microkelvin Imager \citep[AMI;][]{Staley2013MNRAS} who have demonstrated typical response times for follow-up of \emph{Swift}-BAT triggers of order 5 mins.

The ASKAP\footnote{Australian SKA Pathfinder \citep[ASKAP;][]{Johnston2007PASA}} Survey for Variables and Slow Transients \citep[VAST;][]{Murphy2013PASAa} is a Survey Science program that will operate in Western Australia. Phased array feed technology on each of 36 antennas 12-m diameter will allow an instantaneous field of view of 30 deg$^{2}$ at 1.4 GHz, with a resolution of 10-30 arc sec at 300 MHz bandwidth, and a frequency range of (0.7 -- 1.8) GHz. The VAST transient detection pipeline will operate on an imaging cadence of 5–-10 seconds at its fastest down to a cadence of minutes depending on available supercomputing resources. Longer cadences of hours to months can also be conducted through repeated observations of selected fields. When commissioning is complete, ToO response timescales could be as short as minutes when operating in fully automated mode. This accompanied by the large FoV could provide good follow-up opportunities for very low-latency triggers.

In the Netherlands, Apertif \citep[e.g.][]{Verheijen2008,Oosterloo2010} will significantly increase the field of view of the Westerbork Synthesis Radio Telescope and will operate at 1.4 GHz. In standard observing modes, the field of view will be 8 deg$^2$, but has the potential to be extended to 100 deg$^2$ in a ``fly's eye'' mode \citep{Oosterloo2010}.

The Murchison Widefield Array \citep[MWA;][]{Tingay2013PASA} is a Western Australia based low-frequency radio telescope operating between 80 -- 300 MHz. 
It consists of 2048 dual-polarization dipole antennas arranged as 128 tiles. The array has no moving parts, and pointing is achieved by electronically steering\footnote{Telescope functions including pointing are performed by electronic manipulation of dipole signals.}. This provides the ability to retroactively point, at a reduced sensitivity, to a location on the sky in response to a GW event trigger. In addition, implementation of a VOEvent triggering pipeline will reduce the response of this instrument to 10s of seconds \citep{Comm_MWA_2015}. 

The MWA's large FoV -- 600 deg$^{2}$ at 150 MHz defined by the FWHM of the primary beam -- can be extended to 1600 deg$^{2}$ by taking the entire central lobe of the primary beam. At lower frequencies, the FoV scales proportionally to the square of the wavelength, so at 100Hz one can obtain a FoV $\sim 3600$, almost 10\% of the celestial sphere at a single pointing \citep{Kaplan_inprep}. Given a sufficient event rate, the MWA could be pointed at a single patch of sky guaranteeing that 10\% of GW triggers would be captured with zero latency. This improved FoV can be augmented by the MWA Voltage Capture System \citep[VCS][]{Tremblay2015} which would allow data to be simultaneously collected in a 2s cadence imaging mode and in continuous voltage data collection mode. The VCS can be used as a deep disk-based ring buffer with a 1 hr duration. On receipt of a GW trigger, voltage data around the on-source window could be siphoned off for post-trigger analysis.

The Low Frequency Array \citep[LOFAR;][]{vanHaarlem2013AA} is an interferometric array of dipole antenna stations distributed throughout the Netherlands and several countries in Europe. LOFAR uses two distinct antennae types: Low Band Antennae (LBA) operating between 10 -- 90 MHz and High Band Antennae (HBA) operating between 110 -- 250 MHz. The antennae are organised in aperture array stations, with initial beam forming conducted on the station level. These stations have no moving parts and electronic beam-forming techniques allow for rapid re-pointing of the telescope. Due to strategies required to manage the high data rates from LOFAR, the FoV is constrained during the initial processing stages; it is highly dependent upon the station configurations chosen and the observing frequency. Larger FoVs can be attained by using shorter baseline configurations but with a loss of sensitivity and resolution. The maximum FoV (whole observable sky when operating with the LBA) will be monitored in a 24-7 commensal observing mode by (AARTFAAC)\footnote{
 AARTFAAC will have a FoV of 10000 deg$^2$ covering 25\% of the sky. See \url{http://www.aartfaac.org/}} \citep[the Amsterdam-ASTRON Radio Transients Facility and Analysis Centre;][]{Prasad2014} 
At the time of writing, LOFAR still requires manual intervention for external triggers. However, the expectation is that VOEvent triggering will be enabled in time for the first LIGO science runs, reduce the latency to less than a minute. In addition, LOFAR are also developing rapid response strategies based around transient buffer boards \citep[][]{vanHaarlem2013AA} which, similar to the VCS used by MWA, can store small amount of the full observed data for later analysis.

In the longer term, the Square Kilometre Array\footnote{\url{https://www.skatelescope.org/}} (SKA) will be provide large fields of view and increased sensitivity, with construction scheduled to start in 2018. SKA phase 1 will comprise of two instruments SKA-low, operating at 50 -- 350 MHz, and SKA-mid, operating at 350 MHz -- 14 GHz.

\section{A review of EM emissions from BNS mergers}

\label{sec_astro}
A number of associated EM counterparts have been proposed for BNS mergers, many of which have already been presented in Figure 2. We will begin this section by expanding on the multi-messenger pathways and end products provided in Figure 2 which could produce observable EM counterparts for binary coalescences. We will discuss the various scientific breakthroughs that could be achieved in the advanced detector era through low-latency follow-ups and discuss the type of EM observational bands and facilities that could benefit the most from low-latency GW triggers.

Although the scope is wide, the relative brightness of the associated EM signals is dependent on poorly understood merger physics, in particular that of the post-merger central remnant object. Therefore, the prospects of detecting many of these EM counterparts are quite speculative. To address some of this uncertainty, in the latter parts of this section we use the best available EM data to indicate the possible flux of the possible EM counterparts. We do this in the context of EM telescopes already discussed in section 4 which will be available for follow-up observations during the aLIGO era.

Short GRBs have long been proposed to originate from mergers of compact object binaries \citep{paczynski86,eichler89,paczynski91,narayan92}. Recent broad-band observations have shown mixed host galaxy types, non-detection of supernova associations and offsets of GRB locations from their host galaxies \cite[see][for a review]{berger14}. These findings lend support to compact binary mergers as the progenitors for SGRBs.

Generally the merger of BNSs can produce a number of different products; these can be divided into four categories: \\
\noindent 1) A black hole.\\
2) A hyper-massive NS supported by differential rotation; this is expected to collapse to a BH within 10-100\'ms.\\
3) A stable NS that does not collapse.\\
4) A supra-massive NS supported by rigid rotation; this is expected to collapse to a BH within 10-10000\'s with a typical timescale of hundreds of seconds.\\

If the merger product is of the first two catagories: 1) a black hole or 2) a temporal hyper-massive neutron star, the EM signals could include the SGRB and its afterglow emission \citep{eichler89,rosswog13,gehrels05,barthelmy05a,berger11} and a long-lasting radio afterglow of the ejecta as it interacts with the ambient medium  \citep{nakar11,MetzgerBerger2012ApJ,piran13}.

During a NS-NS merger or NS-BH merger, a small fraction of materials are tidally ripped and dynamically ejected. These ejecta are mostly composed of radioactive elements, which would heat the ejecta via decay of r-process nuclei, powering an UV/optical/IR transient \citep{lipaczynski98,kulkarni05,metzger10,barnes13} that has been termed a \emph{kilonova}.  Recent optical and near-infrared observations of a kilonova associated with GRB 130603B have provided support for this scenario and thus a compact object merger origin of SGRBs \citep{Tanvir2013Natur,Berger2013ApJ}.

It is important to note that the SGRB and its afterglow are relativistic and collimated \citep{burrows06,depasquale10}, so only detectable in preferred directions. Alternatively, the kilonova ejecta and its radio afterglow are essentially isotropic and also non-relativistic (due to the heavy mass loading) and therefore can be detected from any direction if the flux is high enough.

The latter two categories, 3) and 4) can both lead to a magnetar as the merger product; this could occur through a small enough total NS mass in combination with a stiff equation of state of nuclear matter. This could be a stable magnetar (for case 3) or a supra-massive millisecond magneter (case 4), which would very likely be strongly magnetized and rapidly spinning \citep[e.g.][]{Dai1998,Dai2006Sci,fanxu06,gaofan06,Yu2007,Zhang2013ApJ,giacomazzo13,Hotokezaka2013,Shibata2014,Lasky2014PhRvD,Ravi2014MNRAS}.

Plateaus in the X-ray afterglow light curves of about 60\% short GRBs have been attributed to such millisecond magnetar central engines \citep{Rowlinson2010,Rowlinson2013MNRAS,Lu2015}. The plateau emission of these SGRBs is interpreted as internal dissipation of the magnetar wind due to the dipole spindown of the magnetar \citep{Zhang2001,Troja07,Metzger2011}. The magnetic spin-down of the NS remnant provides an additional source of sustained energy injection, which greatly enriches the EM signals. A jet may be still launched via accretion from a disk to power the SGRB emission \citep{metzger08,dessart09,lee09,fernandez13}. In the SGRB afterglow phase, some observed features, including the extended prompt emission \citep{norris06,metzger08}, X-ray flares \citep{barthelmy05b,campana06}, and more importantly, plateaus observed in the X-ray are consistent with being powered by a millisecond magnetar central engine.

Outside the solid angle extended by the SGRB jets, several potentially observable astrophysical processes have been predicted. First, self-dissipation of the magnetar wind could power a bright early isotropic X-ray afterglow \citep{Zhang2013ApJ}. Next, heating by the magnetar wind (expected to be more significant than the radioactive decay) could significantly enhance the kilonova emission component \citep{yu13, metzgerpiro14}. Finally, the magnetar wind could accelerate the ejecta to a mildly or even moderately relativistic speed, giving rise to a strong broad-band afterglow emission upon interaction with the ambient medium \citep{Gao2013}.

Finally, a remarkable new class of extremely bright and short-duration (millisecond timescale) transients called \emph{Fast Radio Bursts} (FRBs) could be a product of BNS merger \citep{Thornton2013Sci,Lorimer2013MNRAS}. An FRB could be emitted through a BNS merger in two possible ways:\\
(1) during the BNS merger a coherent radio emission could be produced through magnetic braking as the magnetic fields of the NSs are synchronized in binary rotation at the time of coalescence \citep{Totani2013PASJ}; \\
(2) if a supramassive NS is the merger product, it would collapse into a black hole after loosing the centrifugal support and emit an FRB \citep{Zhang2014ApJ} due to the ejection of the magnetosphere as it implodes into the black hole \citep{Falcke_Rezzolla2014AA}.

The latter scenario is evidenced by an abrupt drop in the X-ray plateau emissions of some long and short GRBs. Within this latter scenario, an FRB along the jet direction is expected after an BNS merger event that forms the supra-massive NS.

In the remainder of this section, we will discuss each of the potential electromagnetic counterparts in more depth and compare the expected emission to the capabilities of the instruments described in Section 4. First, in Section 5.1, we outline the methods used to quantify the typical emission we might expect. Section 5.2 considers the on-axis emission expected from our understanding of SGRBs and Section 5.3 discusses the isotropic kilonova counterpart. In Section 5.4 we describe the impact of changing the central engine formed by the BNS merger from a black hole to a magnetar on the standard emission as discussed in Sections 5.2 and 5.3. Finally, in Section 5.5, we consider the possibility of early-time FRBs associated with a BNS merger and how low-latency triggers and existing instrumentation can constrain these predictions.

\section{Assessing prompt EM follow-ups from low-latency GW triggers}

\label{sec_astro_flux}
\subsection{Extrapolation of fluxes for observed EM counterparts}
In this section we explore the benefits of low-latency GW follow-up observations by extrapolating observed emissions from known sources associated with BNS mergers to the observational range of ground based GW detectors. SGRBs are the most likely observed population to be associated with BNS. Therefore, we can estimate the typical fluxes that could be expected at different energies by extrapolating from known SGRBs associated with BNS at a distance of 200 Mpc. The results are shown in Table \ref{table_fluxes} however, due to the nature of the SGRB observations, these numbers focus on the on-axis emissions. These values are representative of the expected on-axis signals but we do not give the full range of expected fluxes; this requires detailed modelling, beyond the scope of this paper. Where possible, we utilise GRB 130603B as it is the only unambiguously SGRB associated with a BNS merger with a spectroscopic redshift from the optical afterglow \citep{Barthelmy2013,Fong2014,deUgartePostigo2014}. GRB 130603B does not exhibit the all of the emission combinations we consider and, in those cases, we use other SGRBs with reasonable host galaxy redshift constraints.
\begin{table*}
   \caption{The typical expected emission from an on-axis compact binary merger at a distance of 200 Mpc for given observing latencies (\texttt{-} = no emission expected, \texttt{?} = possible emission, refer to text for more details). We provide different predictions for the X-ray emission representing the range of different lightcurves observed from SGRBs \citep[interpreted as different energy injection signatures;][]{Rowlinson2013MNRAS} and separate the SGRBs with bright and faint optical counterparts.}
   \centering
   \begin{tabular}{c|ccccccc} 
    \hline
     \hline
  Wavelength & \multicolumn{5}{c}{EM follow-up time}   \\
   \hline
             & 0 s & 2min & 10 min & 1 hour & 9 hours \\
  \hline \hline
  VHE $\gamma$-ray ($>$100 MeV, ph~cm$^{-2}$s$^{-1}$)        & $1.1\times10^{1}$ & $1.2\times10^{-2}$ & $<1.7\times10^{-3}$ & \texttt{-} & \texttt{-}  \\
  $\gamma$-ray (15 -- 150 keV, ph~cm$^{-2}$s$^{-1}$)       & $3.2\times10^{2}$ & \texttt{?} & \texttt{?}  & \texttt{-} & \texttt{-} \\
  X-ray - no energy injection (0.3 -- 10 keV, erg~cm$^{-2}$s$^{-1}$) & \texttt{-} & $7\times10^{-12}$ & $4\times10^{-12}$ & $2\times10^{-12}$ & $9\times10^{-13}$ \\
  X-ray - unstable magnetar (0.3 -- 10 keV, erg~cm$^{-2}$s$^{-1}$)   & \texttt{-} & $3\times10^{-9}$ & $6\times10^{-11}$ & $5\times10^{-13}$ & $1\times10^{-15}$ \\
  X-ray - stable magnetar (0.3 -- 10 keV, erg~cm$^{-2}$s$^{-1}$)   & \texttt{-} & $1\times10^{-8}$ & $6\times10^{-9}$ & $2\times10^{-9}$ & $8\times10^{-11}$ \\
  Optical - bright (R band, mag)                  & \texttt{?} & 17.0 & 16.8 & 16.5 & 16.9 \\
  Optical - faint (R band, mag)                   & \texttt{?} & 17.9 & 19.0 & 20.2 & 21.8 \\
  Radio (6.7 GHz, mJy)                            & \texttt{?} & \texttt{?} & \texttt{?}  & \texttt{?}  & 11   \\
  \hline
   \end{tabular}
   \label{table_fluxes}
\end{table*}

The X-ray, $\gamma$-ray and radio flux estimates are obtained by extrapolating to the required observation time from broken power law fits to the lightcurves of the GRBs, which are converted into rest-frame luminosities using a k-correction and the observed spectral slope \citep{Bloom2001}. The observed flux at a distance of 200 Mpc is calculated by converting the rest-frame luminosities using an inverted k-correction for the new distance. The standard $\gamma$-ray emission is extrapolated from GRB 130603B \citep{Barthelmy2013}, while the long lasting VHE $\gamma$-ray emission is extrapolated from {\it Fermi} LAT observations of GRB 090510 \citep{DePasquale2010}. In Sections 5.2.2 and 5.4, we consider three different X-ray counterparts and use three SGRBs with reasonable redshift constraints to quantify the typical fluxes given in Table \ref{table_fluxes}, namely: GRB 050509B for a SGRB without energy injection in the X-ray afterglow \citep{gehrels05,Tunnicliffe2014MNRAS}, GRB 130603B for a long lived afterglow showing a clear signature of energy injection associated with a stable magnetar central engine \citep{Fan2013PhRvD,deUgartePostigo2014,Fong2014} and GRB 080905A for an X-ray lightcurve with the signature of energy injection from a short-lived central engine, such as an unstable magnetar central engine \citep{Rowlinson2010MNRASb,Rowlinson2013MNRAS}. The radio fluxes given in Table \ref{table_fluxes} are extrapolated from GRB 130603B \citep{Fong2014}.

The optical magnitudes are determined at 200 Mpc using the standard formalism\footnote{The standard formula is, $m - M = 5 \log D - 5$, with $m$ the apparent magnitude, $M$ is the absolute magnitude and $D$ is the luminosity distance in parsecs}. The optical afterglows are considered in Section 5.2.2; we typically observe very faint and rapidly fading optical afterglows to SGRBs and we represent these counterparts in Table \ref{table_fluxes} by extrapolating the optical counterpart to GRB 080905A \citep{Rowlinson2010MNRASb}. GRB 130603B had an unusually bright optical counterpart and we use this GRB to represent the bright optical afterglow in Table \ref{table_fluxes}. Additionally, in Section 5.3, we consider the isotropic kilonova counterpart to GRB 130603B which would have an observed magnitude of $\sim$21 in the J band, but undetected in the R-band to $\gtrsim$23.7, at a distance of 200 Mpc.

In the following Sections, by combining the results from Tables \ref{table_high_energy}-\ref{table_fluxes} we discuss some possible observational opportunities and implications provided through a low-latency window.
\subsection{Short duration GRBs}
\subsubsection{Probing the progenitor models of Short GRBs}
\label{sgrb_prompt}
Coincident EM/GW observations of SGRBs can prove without doubt that BNS mergers are the progenitors behind these bursts. It is quite possible that 1-2 coincident events a year may be accessible at full aLIGO/AdV sensitivity using GRB satellites \citep{MetzgerBerger2012ApJ,Siellez2014MNRAS,Regimbau2015ApJ}. Such a verification could be provided by a wide FoV instrument such as \emph{Fermi}. Low-latency follow-ups could also enable {\em multi-wavelength} measurements of the prompt emission allowing the underlying central engine and emission mechanism to be probed \citep{Elliott2014A&A}.

A particular obstacle faced by \emph{Swift}-led observations of SGRBs is that the observed $\gamma$-ray emissions typically have observed durations shorter than 2 seconds. Therefore even a GW trigger with a latency as short as 10\,s wold not be sufficient to allow optical telescopes to recover emission during the prompt phase. However, a GW trigger sent out during the inspiral phase would allow multi-wavelength telescopes be on target before the SGRB. Such a scenario could potentially provide a breakthrough in our understanding SGRB prompt emission physics.

In fact a number of foreseeable opportunities to study SGRB physics exist that could be plundered through low-latency follow-ups. Within the GRB theoretical framework, panchromatic emission during the prompt phase and early afterglow phase is predicted, including the emission from the prompt emission site through internal shocks \citep[e.g.][]{Meszaros_Rees1993ApJ}. Indeed, for long duration GRBs, panchromatic emission has been observed in a handful of \emph{Swift} GRBs \citep[e.g.][]{page07,zheng12} thanks to their long durations or the existence of a precursor that triggered the \emph{Swift} BAT. In particular, prompt optical flashes have been observed in a handful of long GRBs: e.g. the \emph{Naked eye burst} GRB 080319B \citep{Racusin2008Natur}, GRB 990123 \citep{Akerlof1999Natur}, GRB 041219A \citep{Vestrand2005Natur}, and GRB 130427B \citep{Vestrand2014Sci}.
However, so far no emission outside the $\gamma$-ray band has been detected during the prompt emission phase of SGRBs. Low-latency GW triggers 10s of seconds before merger could make this possible for the first time. As noted in section 4.2, such observations will become more feasible once  the error region is decreased through an expanded GW network in the latter part of the advanced GW detector era. For a particularly bright or close counterpart, the wide FoV and fast response of an instrument such as GWAC could prove to be highly valuable.

The high energy reach of Fermi has allowed access to photons with energies in access of 10 GeV in both long and SGRBs. The fact that Fermi LAT discovered a photon of energy 31 GeV during the prompt phase of SGRB 090510 \citep{Ackermann_2010ApJ} is quite significant for low-latency follow-ups at VHE. In Table \ref{table_fluxes}, we provide the expected VHE ($>100$ GeV) $\gamma$-ray flux 
and $\gamma$-ray flux 
for a putative SGRB at 200 Mpc coincident with a GW source. A low latency trigger GW trigger would alert VHE and $\gamma$-ray detectors to search for prompt EM counterparts. In the $\gamma$-ray band the wide FoV of an instrument such as Fermi would be advantageous for such rapid follow-ups. At higher energies, instruments such as CTA could provide follow-ups in wide-field mode \citep{Bartos2014MNRAS}; such observations would be very rich in high energy photons and could provide valuable clues on the prompt emission mechanisms of SGRBs \citep{Inoue2013252,Bartos2014MNRAS}.

Additionally, we note that there are a small population of SGRBs that exhibit extended $\gamma$-ray emission for hundreds of seconds following the initial short burst \citep[e.g.][]{norris06}. This sub-population of SGRBs are also believed to originate from compact binary mergers, and hence we might expect to observe this extended emission in some of the triggers even if the bright short-hard emission is missed by the $\gamma$-ray detectors due to the inadequate time for slew from the GW trigger. The possible response times provided for CTA in Table 3 suggest that VHE follow ups of these emissions are quite possible. Such observations could potentially unravel the exact nature of this extended activity \citep{Inoue2013252,Veres_2014ApJ}.

\subsubsection{Capturing the afterglow and late-time central engine activity of SGRBs}
\label{afterglow}
By comparing the expected X-ray counterparts given in Table \ref{table_fluxes} to the instruments described in table \ref{table_high_energy}, we show that existing X-ray instruments such as \emph{Swift} XRT or {\it SVOM} MXT could detect the early X-ray afterglow for the SGRBs even without energy injection. Unfortunately, if the X-ray observations start with a latency longer than $\sim$10 minutes following the BNS merger, or they require a long time to tile out regions of the sky due to having a small FoV, we are likely to miss the X-ray counterparts for SGRBs without prolonged energy injection in their lightcurves (such as those described in Section 5.4).

We consider both bright and faint optical afterglows in Table \ref{table_fluxes} and note all of the optical instruments would be capable of observing the earliest onset (or the rise) of the optical afterglow assuming the 17th magnitude extrapolation based on a burst like GRB 130603B. However, GRB 130603B was unusually bright at early times, with a likely optical plateau phase, so the optical emission may have been contaminated with energy injection from the central engine \citep{deUgartePostigo2014}. Hence, these values are optimistic limits. Most SGRBs typically have a fainter counterpart, such as GRB 080905A, so will require deeper optical observations. The short optical reach and short exposure times of Pan-STARRS and ZTF suggest these instruments could be valuable in such endeavors. The larger FoV of ZTF would be more suited to searching the error regions from earlier GW triggers and could facilitate a co-ordinated search strategy with an instrument such as Pan-STARRS capable of deeper searches but more limited by FoV.

In the radio domain, there have been 3 detections of radio emission associated with SGRBs (GRB 050724,GRB 051221A and GRB 130603B) as well as a forth case, GRB 120804A, due to host galaxy emission. In Table \ref{table_fluxes}, we provide fluxes based on the radio counterpart of GRB 130603B. All of these detections have been at late times ($>$ 7 hours) in the high frequency domain ($> 4.9$ GHz) with fluxes around the mJy level. These fluxes could be observed by instruments such as ASKAP or APERTIF. The peak in the radio afterglow would be expected to occur on typical timescales of days - years. At low radio frequencies, we may also expect to observe the afterglow on months - years timescales using facilities such as MWA and LOFAR although this emission has not been observed to date for a SGRB.

During the early afterglow phase, a short-lived backward propagating \emph{reverse shock} component can occur once the shocked matter encounters the interstellar medium. This may dominate the optical and radio band under certain conditions, e.g. if the reverse shock region is more magnetized than the forward shock region \citep{Zhang03_ApJ}.

On short timescales, an early radio flash ($<$ 1 day) originating from a reverse shock may be expected. Reverse shocks have been observed for a number of long GRBs, peaking a few hours after the GRB, \citep[e.g. GRB 130427A][]{Anderson2014MNRAS,Perley2014ApJ} and may already have been observed in some SGRB radio afterglows \citep[e.g. a radio excess was potentially observed at early times for GRB 130603B;][]{Gompertz2015}. Early radio observations, within the first hours of the BNS merger, capturing the rise of the reverse shock at different radio frequencies would allow detailed modeling of this mechanism and would be a valuable probe of the jet models, macrophysics within the jet and the surrounding environment of the GRB \citep[as conducted for the long GRB 130427A by, e.g.,][]{vanderhorst2014}. With its expected rapid triggering capabilities, ASKAP has the capability to detect an early reverse shock.
\subsection{Kilonova}
\label{kilonova}
The peak frequency and brightness of a kilonova critically depend on the amount of ejected materials, and more importantly, the opacity of the ejecta. Early calculations adopted a relatively small opacity \citep[e.g.][]{lipaczynski98,metzger10,Bauswein2013ApJ}, and predicted an optical transient with a peak flux about 1000 times of nova luminosity (and hence, the term ``kilonova'' is coined) \citep{metzger10}. \cite{barnes13} pointed out that the existence of heavy elements, especially lanthanides, greatly increase the opacity, so that the kilonova signature would peak at a later time and in a softer frequency, typically in IR (see also \citealt{Tanaka2013}).

The properties of the kilonova depend on the properties of the post-merger compact object. The calculations of \cite{barnes13} are relevant for the case of prompt formation of a black hole. If the merger product is a supra-massive neutron star (sustained by rigid rotation for 100s of seconds or even longer, SMNS) or even a hyper-massive neutron star (sustained by differential rotation for 100s of milliseconds, HMNS), neutrinos from the NS would raise the electron fraction and reduce the opacity, so that the kilonova signature is bluer and earlier \citep{Metzger_2014}.

As discussed earlier, an observational candidate for a kilonova was claimed for SGRB 130603B \citep{Tanvir2013Natur,Berger2013ApJ} with one HST excess data point in the near IR band 9 days after the SGRB. As shown in Section 5.1, this kilonova is equivalent to a J band magnitude of 21 at 200 Mpc and an R band magnitude of $\gtrsim$23.7. None of the optical telescopes given in Table 4 have near IR capabilities and are not expected to detect an R-band counterpart at this observation time. Detailed modelling will be required to estimate the R band fluxes and peak timescales for these instruments \citep[see the work by][]{Tanaka2013,barnes13}. Besides this one case, two other candidates have been reported. One is an excess around 13.6 days in the HST F814W-band following GRB 060614 \citep{yang15}; another is a significant optical rebrightening feature around 1 day following GRB 080503 \citep{perley09,Gao2015ApJ}. More data are needed to reveal the diverse kilonova phenomenology. A low latency trigger would facilitate broad-band follow up observations to catch the kilonova signals, especially the early ones due to the small opacity in the ejecta. This will be a challenge due to the lack of sensitive instruments in the NIR band with FoVs large enough to survey the error regions \citep{Bartos_JWT_Kilo_2015}. As discussed in the next section, an associated X-ray emission could be possible; this would allow a wide field instrument such as ISS-Lobster to narrow down the search area.

\subsection{Magnetar formation}
\label{Magnetar formation}
The X-ray plateaus attributed to a millisecond magnetar central engine are observed up to 100-10000s after the bursts. The estimates in Table \ref{table_fluxes} show that if a stable magnetar is produced, GW triggers with a moderate latency could easily allow an X-ray instrument such as \emph{Swift} to detect the plateau emission. However, should an unstable magnetar be produced, a GW trigger with a 40s latency would be required enable an EM follow up within a few 100s. Such observations of the platau phase in combination with GW information could provide important clues of the NS equation of state \citep{Lasky2014PhRvD,Lu2015}.

In addition to the X-ray plateaus, X-ray flares observed in the first few hundred seconds following short GRBs have also been interpreted as resulting from magnetic activity in a post merger differentially rotating supramassive NS \citep{Dai2006Sci}. Prompt follow ups of the GW signals from NS mergers would provide significant evidence for this formation channel of magnetars.

In the massive proto-magnetar central engine model, the potentially wide-beamed magnetar wind is expected to produce a bright broadband afterglow \citep{Zhang2013ApJ,Gao2013}. This could result in an early afterglow in the optical (possibly as bright as 17th magnitude in R band) and X-ray bands (up to $10^{-8}-10^{-7}$ erg s$^{-1}$ cm$^{-2}$) for a source at $\sim 300$ Mpc \citep{Zhang2013ApJ}.

The existence of a supra-massive NS can also supply continuous heating to the ejecta which can, depending on the spindown parameters of the neutron star,  greatly enhance the kilonova signal \citep{yu13,metzgerpiro14}. The detectability also depends on the observational geometry. For an on-axis binary merger, the signal may be fainter than the SGRB optical afterglow. However, for off-axis mergers, this would be the main UV/optical/IR signal from the event. The radio emission is also expected to peak earlier and brighter in this central engine model \citep[e.g.][]{Gao2013,Metzger2014b,Gompertz2015MNRAS}. These predicted optical and radio counterparts are potentially within the sensitivity and field of view capabilities of many of the optical and radio telescopes described in Tables 4 and 5.

Additionally, recent suggestions that the observed prompt emission does not necessarily signal the start of the X-ray emission could have important consequences for low latency follow-ups. \citet{Rezzolla2014, Ciolfi2014} have postulated that the X-ray emission could start before the observed short hard spike emission of short GRBs. These models predict that X-ray emission from a proto-magnetar could occur 100-1000s prior to the prompt GRB.

It is clear that low latency observations will play a major part in testing these models, but these will require more rapid triggering pipelines than those achievable by facilities such as \emph{Swift} and SVOM-MXT (see Table 3). The employment of a wide FoV X-ray instrument with rapid pointing capabilities such as ISS-Lobster \citep{Camp2013} would be very timely for the advanced GW detector era.
\subsection{Fast Radio Bursts}
\label{section_frbs}
The possibility of associations with FRBs and SGRBs and/or BNS mergers is still highly speculative \footnote{However, some models for coherent radio emission have been proposed the late stages of a compact binary inspiral phase or during their coalescence \citep[e.g.][]{Usov2000A&A}.}. \cite{Thornton2013Sci} has suggested that the event rate of FRBs ($\sim 10^3$ gal$^{-1}$yr$^{-1}$) is inconsistent with GRBs and compact star mergers, but could be in agreement with soft gamma-ray repeaters or core-collapse supernovae.  \cite{Totani2013PASJ} argued that including the effects of cosmological time dilation and merger rate evolution would make the FRB event rate in agreement with the optimistic double NS merger event rate \citep{Abadie2010CQGra}. One caveat is that these models suggest an essentially isotropic emission of FRBs. However, due to the existence of the r-process ejecta launched before the coalescence, the FRB signal may not be able to escape the merger environment in most of the solid angles. \cite{Zhang2014ApJ} acknowledged this and propose that the GRB- or merger-associated FRBs only account for a small fraction of the entire FRB population, and only those mergers whose jet direction coincides the observer's viewing direction (i.e. the short GRB direction) would be associated with FRBs. The majority of FRBs are attributed to those SMNSs whose delay time scale to collapse into a black hole is much longer, and therefore not associated with GRBs or supernovae \citep{Falcke_Rezzolla2014AA}.
This is certainly an area where investments in GW low-latency trigger and prompt radio follow-up observations could provide significant scientific insight.

As discussed earlier, a particular obstacle for short duration events of order a few seconds, is that even a latency as short as 10\,s would not be sufficient to alert a $\gamma$-ray instrument in time to recover the prompt emission. This however can be circumvented in the radio band, by electronically steered radio arrays such as LOFAR, MWA and, in the future, SKA-low. An instrument such as MWA could cover the entire GW error region in less than a minute of receiving a trigger. Although LOFAR typically has a smaller field of view than MWA (this is dependent upon observing configuration and frequency), it is also capable of tiling out large regions of the GW error box using multiple simultaneous pointing directions and can conduct commensal beam-formed observations to probe higher time resolutions. In addition, as MWA and LOFAR operate at low frequencies the signal will be delayed by propagation through the ionised intergalactic/interstellar media. Assuming a dispersion measure of 1000 pc\,cm$^{3}$, the delay could be of order 3 minutes at 150 MHz - sufficient time to point the telescope at the error region. With the voltage capture system on MWA and the transient buffer boards on LOFAR, in combination with the low latency triggers and signal delay time, it will be possible for data to be obtained that can constrain emission from seconds to minutes prior to the merger. SKA-low is expected to have many of the capabilities of LOFAR and MWA but with a much higher sensitivity, making it the ideal future instrument for these triggers. For higher frequency instruments such as ASKAP and APERTIF, the delay would be significantly shorter - of order 8\,s at 700 MHz - requiring much more rapid response times. Once on target, these radio telescopes would ideally continuously monitor the entire field for several hours as the different predicted FRB mechanisms cover a range of timescales, both a few seconds prior to the merger and hours following it, and these observations would also be able to constrain the occurance of other types of coherent radio emission.
\section{Discussion and conclusions}
\label{concl}
\begin{figure*}
\includegraphics[scale = 0.7]{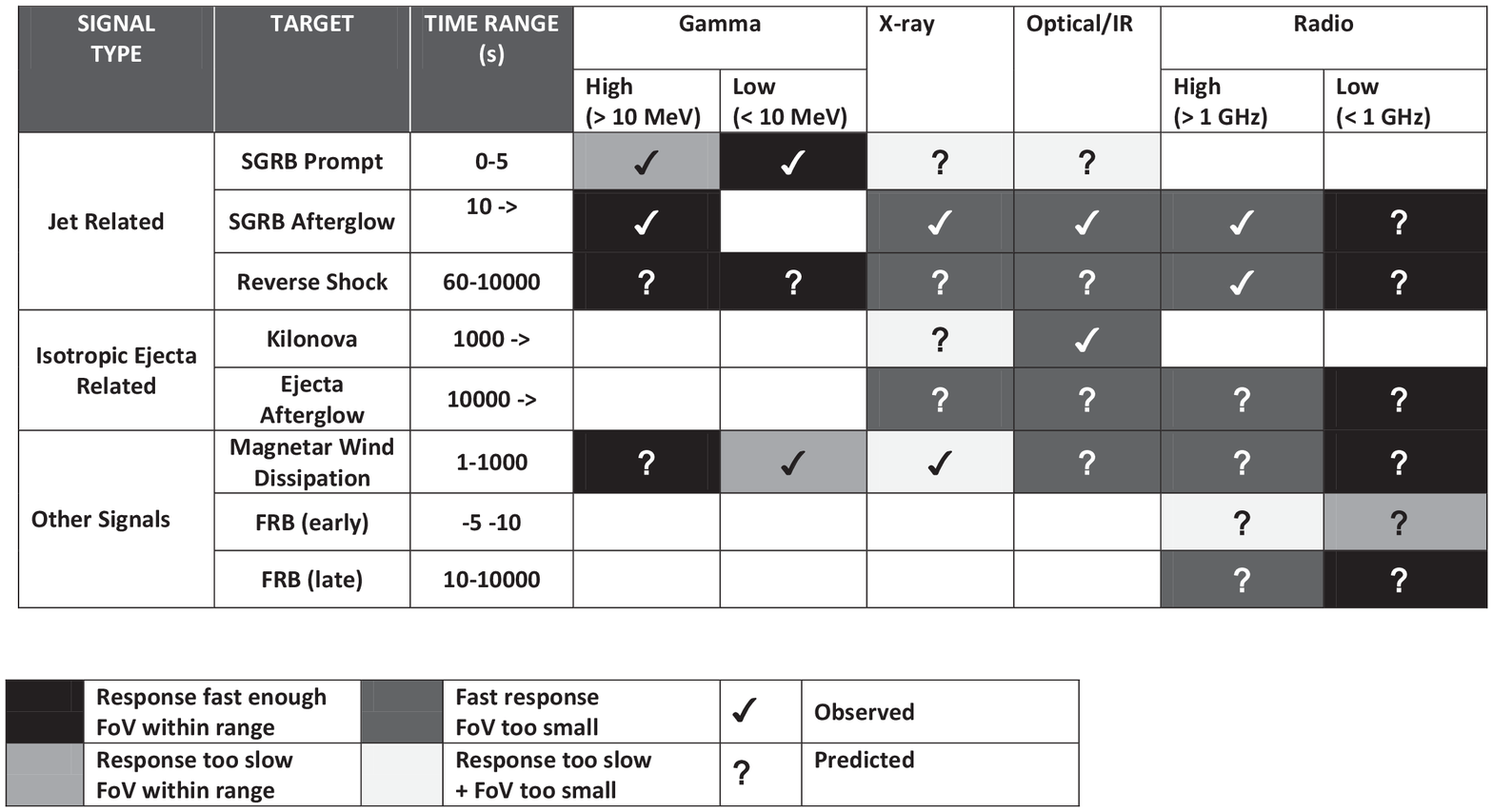}\\
  \caption{Some of the multi-messenger observations that could be achieved through low latency GW triggers. All scenarios assume a detection 40s before merger by an LHV network - corresponding with a 1000 deg$^{2}$ error region - and a 40s GW triggering latency to send out the GW trigger. The signals are grouped by into jetted, isotropic or unknown; the approximate emission time ranges around merger are also given. As shown by the legend, the ticks indicate the EM observations with observational support; the question marks indicate those that have been proposed through theoretical models. All possible observations in the table are given a color code; good news is given by \textbf{dark} shaded cells that indicate there are EM instruments available in the band with sufficiently fast responses, exposure times and FoVs to cover or tile the GW error box within the emission time-range; at the other extreme, \textbf{light} shaded cells indicate that EM follow-up observations will be quite challenging. Between these two extremes there is scope for improvement; for example ToOs could potentially be speeded up or wider FoV instruments could come on line.}
  \label{fig_mm_scenarios1}
\end{figure*}

\begin{figure*}
\includegraphics[scale = 0.7]{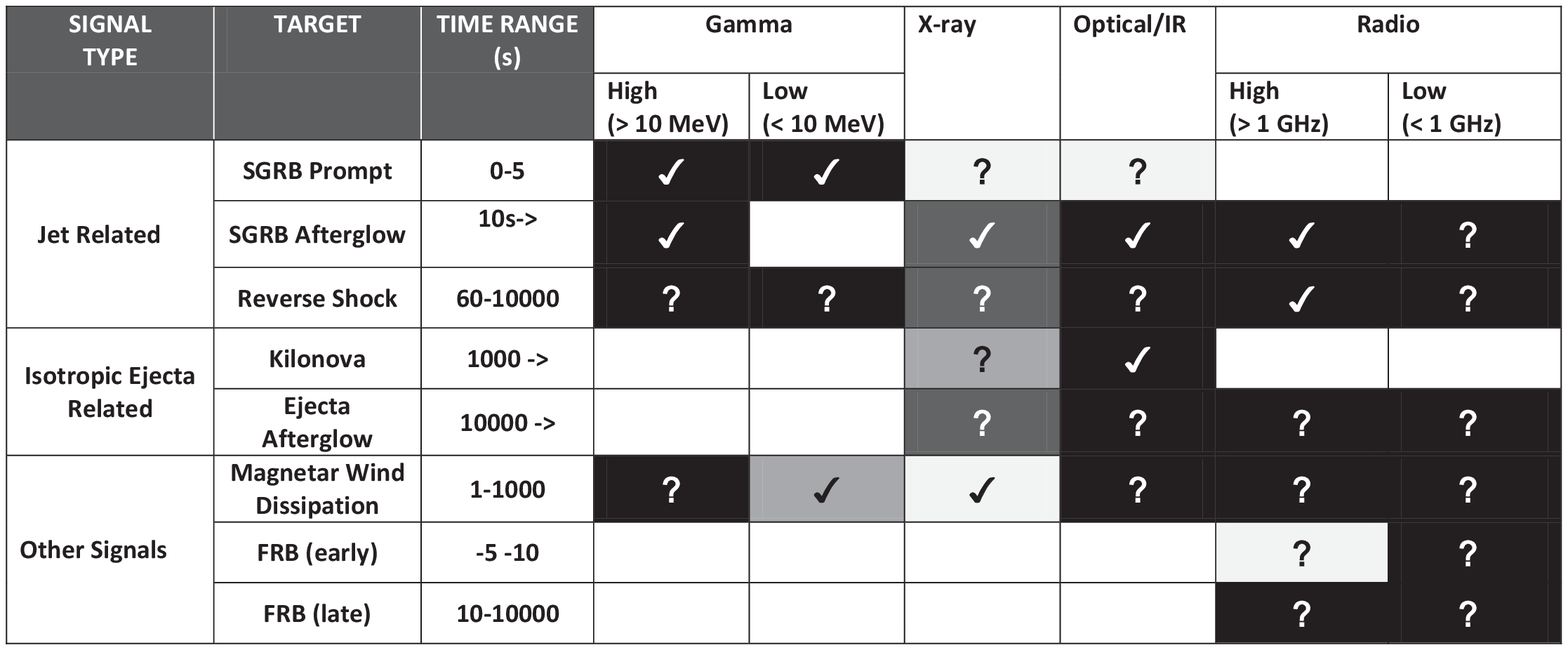}\\
  \caption{As for Figure \ref{fig_mm_scenarios1} but assuming an expanded LHVJIA network -- corresponding with a 182 deg$^{2}$ error region, and a 1s GW triggering latency to send out the GW trigger. This figure shows what could be achieved by EM instruments given a $\sim$40\,s head start to get on-source.}
  \label{fig_mm_scenarios2}
\end{figure*}
The first gravitational wave observation has provided a pathway to the highest energy density events in the Universe and transform our understanding of the behaviour of matter and space-time under the most extreme conditions. Coalescing systems of NSs are widely expected to be the progenitors of short GRBs. Early multiwavelength follow-up observations of these sources would allow one to probe and unravel the mechanisms behind these events beyond the GW regime; the potential for breakthough science is without doubt.

In this study we have simulated the expected localisation error regions corresponding to different times before the merger of a coalescing system of NSs. To examine the most probable low-latency EM observations we have considered the benefit provided by: a) larger networks of GW detectors, corresponding with improved localisation; b) improved GW triggering latencies for sending out GW triggers.

The Tables shown in Figures \ref{fig_mm_scenarios1} and \ref{fig_mm_scenarios2} illustrate some of the multi-messenger observations that could be achieved through low latency GW triggers assuming a detection 40s before merger. We note here that these tables may well be riddled with caveats due to the different operating scenarios and evolution of the different wavelength instruments. Furthermore, it would require very detailed modelling to predict exact fluxes of the different EM emission mechanisms; well beyond the scope of this work. The goal of this study is to highlight the type of breakthrough science that could be realised through the low-latency domain and, using a wide selection of EM instruments with agreements in place with aLIGO/AdV for follow-up observations, determine the most likely avenues for success.

Figure \ref{fig_mm_scenarios1} presents a conservative scenario; that of a LHV network corresponding with a 1000 deg$^{2}$ error region (assuming the 50\% case provided in Table \ref{early_loc}) and a 40s GW triggering latency to send out the alert. At low frequency radio and both high and low-energy $\gamma$-ray, the large FoV and fast response potential show that post-prompt ($>$ 5s) breakthrough observations could be realised.

In the optical, even for Pan-STARRS and ZTF, both large FoV instrument with fast responses, to cover the 1000 deg$^{2}$ error region with an exposure times of 30s would require $>$ (4000;600)s respectively. For longer latency follow-ups, optical instruments available in the latter stages of the aLIGO/AdV era include BlackGEM which has a 40 deg$^{2}$ FoV and a depth of 22 mag and a possible future configuration of GOTO, which could employ a 40-50 deg$^{2}$ instantaneous FoV and sensitivity to 20-21.5 mag \citep{Comm_GOTO_2015}.

To catch the early prompt stage, assuming a a 40s GW triggering latency, even instruments with a large FoV will be challenged; this scenario certainly provides motivation for speeding up the trigger pipelines of both the GW observatories as well as the high energy satellites. We note that for the low-frequency radio, one bonus in regards to EM follow-ups is that the dispersion of coherent radio signals can introduce a significant delay of up to minutes (as discussed in Section \ref{section_frbs}); as the existence of early time counterparts to coalesing BNSs in the radio band are still highly speculative, we have taken a conservative approach and neglected this effect, assuming only the physical time in our estimates.

A more optimistic scenario is presented in Figure \ref{fig_mm_scenarios2} which shows how breakthrough science can be achieved with an almost instantaneous 1s GW triggering latency and a fully expanded advanced detector network, LHVJIA. For the 182 deg$^{2}$ error region the post-prompt ($>$ 5s) situation is much more optimistic in the optical and high-frequency radio bands due to the FoVs of radio instruments with responses $<$mins and wide FoV telescopes such as ZTF that could survey this error region in around 2 mins. In the X-ray, small FoV instruments such as \emph{Swift} are hindered by the large GW error region. Although considerable efforts have been undertaken to remedy this situation through the use a requirement for reconstructed sky locations to overlap with nearby galaxies \citep{Evans2012ApJS}, this coupled with the slow response of high-energy satellites presents a significant bottleneck in a band for which many of the post merger EM emissions are known to exist. It emphasises the importance to have a wide FoV X-ray instrument capable of rapid pointing; one potential instrument that could remedy the situation is the NASA proposed 30 deg$^{2}$ FoV soft X-ray instrument ISS-Lobster \citep{Camp2013}.

For prompt ($<$ 5s) observations the wide FoV of Fermi means this instrument could provide a coincident observation with a GW trigger - this scenario is considered in the low gamma-ray ( $<$ 10 MeV) region of Figures \ref{fig_mm_scenarios1} and \ref{fig_mm_scenarios2}. However, we note the slow response of other satellites operating at the low-$\gamma$ will hinder prompt follow-ups. If such a low GW triggering latency could be achieved and instruments such as \emph{Swift}/Fermi could employ fast triggering pipelines, and a wide FoV X-ray instrument could be employed, low-latency EM followups with a full GW global network would enjoy almost complete coverage for the models we have considered.

We note that the scenarios we have outlined in Figures \ref{fig_mm_scenarios1} and \ref{fig_mm_scenarios2} assume the GW error regions 40s before merger; this choice was to provide the EM facilities with the maximum time for response. However, as shown in Table \ref{early_loc}, closer to the merger both the error regions contract and the percentage of detections sources increases. For sources detected early in the inspiraling stage, subsequent alerts would most likely be sent out as the error box decreases in size; this suggests that a co-ordinated observational and alerting strategy involving both the GW network and both wide and narrower FoV instruments would be optimal procedure for maximising GW detections.

Our results support the argument that much breakthrough science can be achieved though a fast response. This has been evidenced in the post-\emph{Swift} era with prompt follow-up of GRB triggers producing such unexpected phenomena as flares, plateaus, optical flashes and reverse shocks. History therefore suggests that the lasting legacy from the era of prompt multi-wavelength GW follow-ups will undoubtedly be provided by the unexpected surprises we encounter.

\section{ACKNOWLEDGMENTS}
Much of the analysis and discussion in this paper would not have been possible without the valuable input from members of the various EM facilities. In particularly we would like to thank: David Kaplan, Shami Chatterjee, Lisa Harvey-Smith (ASKAP), Dipankar Bhattacharya (Astrosat/CZTI ), Imre Bartos (Columbia U), Paul O’Brien, Jim Hinton and Gavin Rowell (CTA), Peter Jonker (LOFAR), Julie McEnery and Valerie Connaughton (Fermi), Weiming Yuan,  Chao Wu and  Jianyan Wei (NAOC), Fang Yuan (SkyMapper), Duncan Galloway and Danny Steeghs (GOTO) and David Coward (Zadko). We are particularly grateful to Julie McEnery (NASA/GSFC) who contributed some valuable suggestions for section 4.1 and to Paul Laskey (Monash University) who provided some useful comments to improve the manuscript. We acknowledge support from WA's Pawsey Supercomputing Center. EJH acknowledges support from a UWA Research Fellowship. LW is supported by the Australian Research Council. BZ acknowledges NASA NNX14AF85G and NNX15AK85G for partial support. The authors gratefully thank Leo Singer, the assigned reviewer for the LIGO Scientific Collaboration, for conducting a thorough review of the manuscript which included a number of insightful suggestions. Finally, the authors thank the anonymous referee whose valued comments have vastly improved the structure of the manuscript.
\vspace{5mm}

\begin{thebibliography}{}

\bibitem[\protect\citeauthoryear{Aasi et~al.,}{Aasi
  et~al.}{2013}]{ligo_localization}
Aasi J.,  et~al.,, 2013, {P}rospects for {L}ocalization of {G}ravitational
  {W}ave {T}ransients by the {A}dvanced {L}{I}{G}{O} and {A}dvanced {V}irgo
  {O}bservatories

\bibitem[\protect\citeauthoryear{Aasi et~al.,}{Aasi  et~al.}{2015}]{aLIGO_2015}
Aasi J.,  et~al., 2015, Classical and Quantum Gravity, 32, 074001

\bibitem[\protect\citeauthoryear{{Abadie}, {Abbott}, {Abbott}, {Abbott},
  {Abernathy}, {Accadia}, {Acernese}, {Adams}, {Adhikari}, {Affeldt} \& et
  al.}{{Abadie} et~al.}{2012}]{first_low_latency_inspiral}
{Abadie} J.,  {Abbott} B.~P.,  {Abbott} R.,  {Abbott} T.~D.,  {Abernathy} M.,
  {Accadia} T.,  {Acernese} F.,  {Adams} C.,  {Adhikari} R.,  {Affeldt} C.,
  et al. 2012, \aap, 541, A155

\bibitem[\protect\citeauthoryear{{Abadie}, {Abbott}, {Abbott}, {Abernathy},
  {Accadia}, {Acernese}, {Adams}, {Adhikari}, {Ajith}, {Allen} \& et
  al.}{{Abadie} et~al.}{2010}]{Abadie2010}
{Abadie} J.,  {Abbott} B.~P.,  {Abbott} R.,  {Abernathy} M.,  {Accadia} T.,
  {Acernese} F.,  {Adams} C.,  {Adhikari} R.,  {Ajith} P.,  {Allen} B.,    et
  al. 2010, Classical and Quantum Gravity, 27, 173001

\bibitem[\protect\citeauthoryear{Abadie et~al.,}{Abadie  et~al.}{2010}]{abadie}
Abadie J.,  et~al., 2010, Class. Quantum Grav., 27, 173001

\bibitem[\protect\citeauthoryear{{Abadie} et~al.,}{{Abadie}
  et~al.}{2010}]{Abadie2010CQGra}
{Abadie} J.,  et~al., 2010, Classical and Quantum Gravity, 27, 173001

\bibitem[\protect\citeauthoryear{Abadie et~al.,}{Abadie
  et~al.}{2011}]{abadie11}
Abadie J.,  et~al., 2011, ArXiv e-prints 1109.3498

\bibitem[\protect\citeauthoryear{{Abbott}, {Abbott}, {Abbott}, {Abernathy},
  {Acernese}, {Ackley}, {Adams}, {Adams}, {Addesso}, {Adhikari} \& et
  al.}{{Abbott} et~al.}{2016b}]{2016ApJ...818L..22A}
{Abbott} B.~P.,  {Abbott} R.,  {Abbott} T.~D.,  {Abernathy} M.~R.,  {Acernese}
  F.,  {Ackley} K.,  {Adams} C.,  {Adams} T.,  {Addesso} P.,  {Adhikari} R.~X.,
     et al. 2016b, \apjl, 818, L22

\bibitem[\protect\citeauthoryear{{Abbott}, {Abbott}, {Abbott}, {Abernathy},
  {Acernese}, {Ackley}, {Adams}, {Adams}, {Addesso}, {Adhikari} \& et
  al.}{{Abbott} et~al.}{2016a}]{2016PhRvL.116f1102A}
{Abbott} B.~P.,  {Abbott} R.,  {Abbott} T.~D.,  {Abernathy} M.~R.,  {Acernese}
  F.,  {Ackley} K.,  {Adams} C.,  {Adams} T.,  {Addesso} P.,  {Adhikari} R.~X.,
     et al. 2016a, Physical Review Letters, 116, 061102

\bibitem[\protect\citeauthoryear{Acernese et~al.,}{Acernese
  et~al.}{2015}]{AdV_2015}
Acernese F.,  et~al., 2015, Class.Quant.Grav., 32, 024001

\bibitem[\protect\citeauthoryear{{Ackermann}, {Asano}, {Atwood}, {Axelsson},
  {Baldini} et~al.,}{{Ackermann} et~al.}{2010}]{Ackermann_2010ApJ}
{Ackermann} M.,  {Asano} K.,  {Atwood} W.~B.,  {Axelsson} M.,  {Baldini} L.,
  et~al., 2010, \apj, 716, 1178

\bibitem[\protect\citeauthoryear{{Aharonian} et~al.,}{{Aharonian}
  et~al.}{2009}]{Aharonian2009A&A}
{Aharonian} F.,  et~al., 2009, \aap, 495, 505

\bibitem[\protect\citeauthoryear{{Akerlof}, {Balsano}, {Barthelmy}
  et~al.,}{{Akerlof} et~al.}{1999}]{Akerlof1999Natur}
{Akerlof} C.,  {Balsano} R.,  {Barthelmy}   et~al., 1999, \nat, 398, 400

\bibitem[\protect\citeauthoryear{{Anderson}, {van der Horst}, {Staley},
  {Fender}, {Wijers}, {Scaife}, {Rumsey}, {Titterington}, {Rowlinson} \&
  {Saunders}}{{Anderson} et~al.}{2014}]{Anderson2014MNRAS}
{Anderson} G.~E.,  {van der Horst} A.~J.,  {Staley} T.~D.,  {Fender} R.~P.,
  {Wijers} R.~A.~M.~J.,  {Scaife} A.~M.~M.,  {Rumsey} C.,  {Titterington}
  D.~J.,  {Rowlinson} A.,    {Saunders} R.~D.~E.,  2014, \mnras, 440, 2059

\bibitem[\protect\citeauthoryear{Aso, Michimura, Somiya, Ando, Miyakawa,
  Sekiguchi, Tatsumi, Yamamoto, Collaboration et~al.,}{Aso
  et~al.}{2013}]{kagra}
Aso Y.,  Michimura Y.,  Somiya K.,  Ando M.,  Miyakawa O.,  Sekiguchi T.,
  Tatsumi D.,  Yamamoto H.,  Collaboration K.,    et~al., 2013, Physical Review
  D, 88, 043007

\bibitem[\protect\citeauthoryear{{Atwood}, {Abdo}, {Ackermann}, {Althouse},
  {Anderson}, {Axelsson}, {Baldini}, {Ballet}, {Band}, {Barbiellini} \& et
  al.}{{Atwood} et~al.}{2009}]{Atwood2009ApJ}
{Atwood} W.~B.,  {Abdo} A.~A.,  {Ackermann} M.,  {Althouse} W.,  {Anderson} B.,
   {Axelsson} M.,  {Baldini} L.,  {Ballet} J.,  {Band} D.~L.,  {Barbiellini}
  G.,    et al. 2009, \apj, 697, 1071

\bibitem[\protect\citeauthoryear{{Barnes} \& {Kasen}}{{Barnes} \&
  {Kasen}}{2013}]{barnes13}
{Barnes} J.,  {Kasen} D.,  2013, \apj, 775, 18

\bibitem[\protect\citeauthoryear{{Barthelmy}, {Cannizzo}, {Gehrels} \&
  others.}{{Barthelmy} et~al.}{2005}]{barthelmy05b}
{Barthelmy} S.~D.,  {Cannizzo} J.~K.,  {Gehrels} N.,    others. 2005, \apjl,
  635, L133

\bibitem[\protect\citeauthoryear{{Barthelmy}, {Chincarini}, {Burrows}
  et~al.,}{{Barthelmy} et~al.}{2005}]{barthelmy05a}
{Barthelmy} S.~D.,  {Chincarini} G.,  {Burrows} D.~N.,    et~al., 2005, \nat,
  438, 994

\bibitem[\protect\citeauthoryear{{Barthelmy} et~al.,}{{Barthelmy}
  et~al.}{2013}]{Barthelmy2013}
{Barthelmy} S.~D.,  et~al., 2013, GRB Coordinates Network, 14741, 1

\bibitem[\protect\citeauthoryear{{Bartos}, {Huard} \& {Marka}}{{Bartos}
  et~al.}{2015}]{Bartos_JWT_Kilo_2015}
{Bartos} I.,  {Huard} T.~L.,    {Marka} S.,  2015, ArXiv e-prints

\bibitem[\protect\citeauthoryear{{Bartos}, {Veres}, {Nieto}, {Connaughton},
  {Humensky}, {Hurley}, {M{\'a}rka}, {M{\'e}sz{\'a}ros}, {Mukherjee}, {O'Brien}
  \& {Osborne}}{{Bartos} et~al.}{2014}]{Bartos2014MNRAS}
{Bartos} I.,  {Veres} P.,  {Nieto} D.,  {Connaughton} V.,  {Humensky} B.,
  {Hurley} K.,  {M{\'a}rka} S.,  {M{\'e}sz{\'a}ros} P.,  {Mukherjee} R.,
  {O'Brien} P.,    {Osborne} J.~P.,  2014, \mnras, 443, 738

\bibitem[\protect\citeauthoryear{{Bauswein}, {Goriely} \& {Janka}}{{Bauswein}
  et~al.}{2013}]{Bauswein2013ApJ}
{Bauswein} A.,  {Goriely} S.,    {Janka} H.-T.,  2013, \apj, 773, 78

\bibitem[\protect\citeauthoryear{{Bellm}}{{Bellm}}{2014}]{Bellm_2014}
{Bellm} E.,  2014, in {Wozniak} P.~R.,  {Graham} M.~J.,  {Mahabal} A.~A.,
  {Seaman} R.,  eds, The Third Hot-wiring the Transient Universe Workshop {The
  Zwicky Transient Facility}.
pp 27--33

\bibitem[\protect\citeauthoryear{{Berger}}{{Berger}}{2011}]{berger11}
{Berger} E.,  2011, \nar, 55, 1

\bibitem[\protect\citeauthoryear{{Berger}}{{Berger}}{2014}]{berger14}
{Berger} E.,  2014, \araa, 52, 43

\bibitem[\protect\citeauthoryear{{Berger} et~al.,}{{Berger}
  et~al.}{2005}]{Berger2005Natur}
{Berger} E.,  et~al., 2005, \nat, 438, 988

\bibitem[\protect\citeauthoryear{{Berger}, {Fong} \& {Chornock}}{{Berger}
  et~al.}{2013}]{Berger2013ApJ}
{Berger} E.,  {Fong} W.,    {Chornock} R.,  2013, \apjl, 774, L23

\bibitem[\protect\citeauthoryear{{Bloom} et~al.,}{{Bloom}
  et~al.}{2006}]{Bloom2006ApJ}
{Bloom} J.~S.,  et~al., 2006, \apj, 638, 354

\bibitem[\protect\citeauthoryear{{Bloom}, {Frail} \& {Sari}}{{Bloom}
  et~al.}{2001}]{Bloom2001}
{Bloom} J.~S.,  {Frail} D.~A.,    {Sari} R.,  2001, \aj, 121, 2879

\bibitem[\protect\citeauthoryear{{Burrows}, {Grupe}, {Capalbi}, {Panaitescu},
  {Patel}, {Kouveliotou}, {Zhang}, {M{\'e}sz{\'a}ros}, {Chincarini}, {Gehrels}
  \& {Wijers}}{{Burrows} et~al.}{2006}]{burrows06}
{Burrows} D.~N.,  {Grupe} D.,  {Capalbi} M.,  {Panaitescu} A.,  {Patel} S.~K.,
  {Kouveliotou} C.,  {Zhang} B.,  {M{\'e}sz{\'a}ros} P.,  {Chincarini} G.,
  {Gehrels} N.,    {Wijers} R.~A.~M.,  2006, \apj, 653, 468

\bibitem[\protect\citeauthoryear{{Buskulic}, {Virgo Collaboration} \& {LIGO
  Scientific Collaboration}}{{Buskulic} et~al.}{2010}]{mbta}
{Buskulic} D.,  {Virgo Collaboration}   {LIGO Scientific Collaboration} 2010,
  Classical and Quantum Gravity, 27, 194013

\bibitem[\protect\citeauthoryear{{Camp}, {Barthelmy}, {Blackburn}, {Carpenter},
  {Gehrels}, {Kanner}, {Marshall}, {Racusin} \& {Sakamoto}}{{Camp}
  et~al.}{2013}]{Camp2013}
{Camp} J.,  {Barthelmy} S.,  {Blackburn} L.,  {Carpenter} K.~G.,  {Gehrels} N.,
   {Kanner} J.,  {Marshall} F.~E.,  {Racusin} J.~L.,    {Sakamoto} T.,  2013,
  Experimental Astronomy

\bibitem[\protect\citeauthoryear{{Campana}, {Mangano}, {Blustin}, {Brown}
  et~al.,}{{Campana} et~al.}{2006}]{campana06}
{Campana} S.,  {Mangano} V.,  {Blustin} A.~J.,  {Brown} P.,    et~al.,, 2006,
  {The association of GRB 060218 with a supernova and the evolution of the
  shock wave}

\bibitem[\protect\citeauthoryear{{Cannon}, {Cariou}, {Chapman},
  {Crispin-Ortuzar}, {Fotopoulos}, {Frei}, {Hanna}, {Kara}, {Keppel}, {Liao},
  {Privitera}, {Searle}, {Singer} \& {Weinstein}}{{Cannon}
  et~al.}{2012}]{cannon12}
{Cannon} K.,  {Cariou} R.,  {Chapman} A.,  {Crispin-Ortuzar} M.,  {Fotopoulos}
  N.,  {Frei} M.,  {Hanna} C.,  {Kara} E.,  {Keppel} D.,  {Liao} L.,
  {Privitera} S.,  {Searle} A.,  {Singer} L.,    {Weinstein} A.,  2012, \apj,
  748, 136

\bibitem[\protect\citeauthoryear{{Centrella}, {Nissanke} \&
  {Williams}}{{Centrella} et~al.}{2012}]{Centrella2012IAUS}
{Centrella} J.,  {Nissanke} S.,    {Williams} R.,  2012, in {Griffin} E.,
  {Hanisch} R.,   {Seaman} R.,  eds, IAU Symposium Vol.~285 of IAU Symposium,
  {Gravitational Waves and Time-Domain Astronomy}.
pp 191--198

\bibitem[\protect\citeauthoryear{Chu, Wen \& Blair}{Chu et~al.}{2012}]{chuqi12}
Chu Q.,  Wen L.,    Blair D.,  2012, Journal of Physics: Conference Series,
  363, 012023

\bibitem[\protect\citeauthoryear{{Chung}, {Wen}, {Blair}, {Cannon} \&
  {Datta}}{{Chung} et~al.}{2010}]{shinkee10}
{Chung} S.~K.,  {Wen} L.,  {Blair} D.,  {Cannon} K.,    {Datta} A.,  2010,
  Classical and Quantum Gravity, 27, 135009

\bibitem[\protect\citeauthoryear{{Ciolfi} \& {Siegel}}{{Ciolfi} \&
  {Siegel}}{2014}]{Ciolfi2014}
{Ciolfi} R.,  {Siegel} D.~M.,  2014, ArXiv e-prints

\bibitem[\protect\citeauthoryear{{Coward}, {Branchesi}, {Howell}, {Lasky} \&
  {B{\"o}er}}{{Coward} et~al.}{2014}]{Coward2014}
{Coward} D.~M.,  {Branchesi} M.,  {Howell} E.~J.,  {Lasky} P.~D.,    {B{\"o}er}
  M.,  2014, \mnras, 445, 3575

\bibitem[\protect\citeauthoryear{{Coward}, {Howell}, {Piran}, {Stratta},
  {Branchesi}, {Bromberg}, {Gendre}, {Burman} \& {Guetta}}{{Coward}
  et~al.}{2012}]{CowardHowellPiran_2012}
{Coward} D.~M.,  {Howell} E.~J.,  {Piran} T.,  {Stratta} G.,  {Branchesi} M.,
  {Bromberg} O.,  {Gendre} B.,  {Burman} R.~R.,    {Guetta} D.,  2012, \mnras,
  425, 2668

\bibitem[\protect\citeauthoryear{{Coward}, {Todd}, {Vaalsta}, {Laas-Bourez},
  {Klotz}, {Imerito}, {Yan}, {Luckas}, {Fletcher}, {Zadnik}, {Burman}, {Blair},
  {Zadko}, {Bo{\"e}r}, {Thierry}, {Howell}, {Gordon}, {Ahmat}, {Moore} \&
  {Frost}}{{Coward} et~al.}{2010}]{Coward2010PASA}
{Coward} D.~M.,  {Todd} M.,  {Vaalsta} T.~P.,  {Laas-Bourez} M.,  {Klotz} A.,
  {Imerito} A.,  {Yan} L.,  {Luckas} P.,  {Fletcher} A.~B.,  {Zadnik} M.~G.,
  {Burman} R.~R.,  {Blair} D.~G.,  {Zadko} J.,  {Bo{\"e}r} M.,  {Thierry} P.,
  {Howell} E.~J.,  {Gordon} S.,  {Ahmat} A.,  {Moore} J.~A.,    {Frost} K.,
  2010, \pasa, 27, 331

\bibitem[\protect\citeauthoryear{{Cowperthwaite} \& {Berger}}{{Cowperthwaite}
  \& {Berger}}{2015}]{cowperthwaite2015}
{Cowperthwaite} P.~S.,  {Berger} E.,  2015, ArXiv e-prints

\bibitem[\protect\citeauthoryear{Cramer}{Cramer}{1946}]{cramer}
Cramer H.,  1946, Mathematical Methods of Statistics.
Princeton University Press

\bibitem[\protect\citeauthoryear{{Dai} \& {Lu}}{{Dai} \& {Lu}}{1998}]{Dai1998}
{Dai} Z.~G.,  {Lu} T.,  1998, \aap, 333, L87

\bibitem[\protect\citeauthoryear{{Dai}, {Wang}, {Wu} \& {Zhang}}{{Dai}
  et~al.}{2006}]{Dai2006Sci}
{Dai} Z.~G.,  {Wang} X.~Y.,  {Wu} X.~F.,    {Zhang} B.,  2006, Science, 311,
  1127

\bibitem[\protect\citeauthoryear{{De Pasquale} et~al.,}{{De Pasquale}
  et~al.}{2010}]{DePasquale2010}
{De Pasquale} M.,  et~al., 2010, \apjl, 709, L146

\bibitem[\protect\citeauthoryear{{De Pasquale}, {Schady}, {Kuin}, {Page},
  {Curran}, {Zane}, {Oates}, {Holland}, {Breeveld} \& {Hoversten}}{{De
  Pasquale} et~al.}{2010}]{depasquale10}
{De Pasquale} M.,  {Schady} P.,  {Kuin} N.~P.~M.,  {Page} M.~J.,  {Curran}
  P.~A.,  {Zane} S.,  {Oates} S.~R.,  {Holland} S.~T.,  {Breeveld} A.~A.,
  {Hoversten} E.~A. e.~a.,  2010, \apjl, 709, L146

\bibitem[\protect\citeauthoryear{{de Ugarte Postigo} et~al.,}{{de Ugarte
  Postigo}  et~al.}{2014}]{deUgartePostigo2014}
{de Ugarte Postigo} A.,  et~al., 2014, \aap, 563, A62

\bibitem[\protect\citeauthoryear{{Dessart}, {Ott}, {Burrows}, {Rosswog} \&
  {Livne}}{{Dessart} et~al.}{2009}]{dessart09}
{Dessart} L.,  {Ott} C.~D.,  {Burrows} A.,  {Rosswog} S.,    {Livne} E.,  2009,
  \apj, 690, 1681

\bibitem[\protect\citeauthoryear{{Doro}, {Conrad}, {Emmanoulopoulos}
  et~al.,}{{Doro} et~al.}{2013}]{Doro2013APh}
{Doro} M.,  {Conrad} J.,  {Emmanoulopoulos} D.,    et~al., 2013, Astroparticle
  Physics, 43, 189

\bibitem[\protect\citeauthoryear{{Dubus} et~al.,}{{Dubus}
  et~al.}{2013}]{Dubus2013APh}
{Dubus} G.,  et~al., 2013, Astroparticle Physics, 43, 317

\bibitem[\protect\citeauthoryear{{Eichler}, {Livio}, {Piran} \&
  {Schramm}}{{Eichler} et~al.}{1989a}]{Eichler1989Natur}
{Eichler} D.,  {Livio} M.,  {Piran} T.,    {Schramm} D.~N.,  1989a, \nat, 340,
  126

\bibitem[\protect\citeauthoryear{{Eichler}, {Livio}, {Piran} \&
  {Schramm}}{{Eichler} et~al.}{1989b}]{eichler89}
{Eichler} D.,  {Livio} M.,  {Piran} T.,    {Schramm} D.~N.,  1989b, \nat, 340,
  126

\bibitem[\protect\citeauthoryear{{Elliott} et~al.,}{{Elliott}
  et~al.}{2014}]{Elliott2014A&A}
{Elliott} J.,  et~al., 2014, \aap, 562, A100

\bibitem[\protect\citeauthoryear{{Evans}, {Fridriksson}, {Gehrels}, {Homan},
  {Osborne}, {Siegel}, {Beardmore}, {Handbauer}, {Gelbord}, {Kennea} \& et
  al.}{{Evans} et~al.}{2012}]{Evans2012ApJS}
{Evans} P.~A.,  {Fridriksson} J.~K.,  {Gehrels} N.,  {Homan} J.,  {Osborne}
  J.~P.,  {Siegel} M.,  {Beardmore} A.,  {Handbauer} P.,  {Gelbord} J.,
  {Kennea} J.~A.,    et al. 2012, \apjs, 203, 28

\bibitem[\protect\citeauthoryear{{Fairhurst}}{{Fairhurst}}{2009}]{fairhurst09}
{Fairhurst} S.,  2009, New Journal of Physics, 11, 123006

\bibitem[\protect\citeauthoryear{Fairhurst}{Fairhurst}{2010}]{fairhurst11}
Fairhurst S., , 2010, {S}ource localization with an advanced gravitational wave
  detector network, Class.Quant.Grav.28:105021,2011

\bibitem[\protect\citeauthoryear{{Falcke} \& {Rezzolla}}{{Falcke} \&
  {Rezzolla}}{2014}]{Falcke_Rezzolla2014AA}
{Falcke} H.,  {Rezzolla} L.,  2014, \aap, 562, A137

\bibitem[\protect\citeauthoryear{Fan, Messenger \& Heng}{Fan
  et~al.}{2014}]{identify_host_galaxies}
Fan X.,  Messenger C.,    Heng I.~S., , 2014, {A} {B}ayesian approach to
  multi-messenger astronomy: {I}dentification of gravitational-wave host
  galaxies

\bibitem[\protect\citeauthoryear{{Fan}, {Wu} \& {Wei}}{{Fan}
  et~al.}{2013}]{Fan2013PhRvD}
{Fan} Y.-Z.,  {Wu} X.-F.,    {Wei} D.-M.,  2013, \prd, 88, 067304

\bibitem[\protect\citeauthoryear{{Fan} \& {Xu}}{{Fan} \& {Xu}}{2006}]{fanxu06}
{Fan} Y.-Z.,  {Xu} D.,  2006, \mnras, 372, L19

\bibitem[\protect\citeauthoryear{{Fern{\'a}ndez} \& {Metzger}}{{Fern{\'a}ndez}
  \& {Metzger}}{2013}]{fernandez13}
{Fern{\'a}ndez} R.,  {Metzger} B.~D.,  2013, \mnras, 435, 502

\bibitem[\protect\citeauthoryear{{Finn}}{{Finn}}{1992}]{finn92}
{Finn} L.~S.,  1992, \prd, 46, 5236

\bibitem[\protect\citeauthoryear{{Fong}, {Berger}, {Metzger}, {Margutti},
  {Chornock}, {Migliori}, {Foley}, {Zauderer}, {Lunnan}, {Laskar}, {Desch},
  {Meech}, {Sonnett}, {Dickey}, {Hedlund} \& {Harding}}{{Fong}
  et~al.}{2014}]{Fong2014}
{Fong} W.,  {Berger} E.,  {Metzger} B.~D.,  {Margutti} R.,  {Chornock} R.,
  {Migliori} G.,  {Foley} R.~J.,  {Zauderer} B.~A.,  {Lunnan} R.,  {Laskar} T.,
   {Desch} S.~J.,  {Meech} K.~J.,  {Sonnett} S.,  {Dickey} C.,  {Hedlund} A.,
   {Harding} P.,  2014, \apj, 780, 118

\bibitem[\protect\citeauthoryear{{Funk}, {Hinton} \& {CTA Consortium}}{{Funk}
  et~al.}{2013}]{Funk2013APh}
{Funk} S.,  {Hinton} J.~A.,    {CTA Consortium} 2013, Astroparticle Physics,
  43, 348

\bibitem[\protect\citeauthoryear{{Gao}, {Ding}, {Wu}, {Dai} \& {Zhang}}{{Gao}
  et~al.}{2015}]{Gao2015ApJ}
{Gao} H.,  {Ding} X.,  {Wu} X.-F.,  {Dai} Z.-G.,    {Zhang} B.,  2015, \apj,
  807, 163

\bibitem[\protect\citeauthoryear{{Gao}, {Ding}, {Wu}, {Zhang} \& {Dai}}{{Gao}
  et~al.}{2013}]{Gao2013}
{Gao} H.,  {Ding} X.,  {Wu} X.-F.,  {Zhang} B.,    {Dai} Z.-G.,  2013, \apj,
  771, 86

\bibitem[\protect\citeauthoryear{{Gao} \& {Fan}}{{Gao} \&
  {Fan}}{2006}]{gaofan06}
{Gao} W.-H.,  {Fan} Y.-Z.,  2006, Chinese Journal of Astronomy and
  Astrophysics, 6, 513

\bibitem[\protect\citeauthoryear{Gehrels et~al.,}{Gehrels
  et~al.}{2004}]{Gehrels_Swift_2004}
Gehrels N.,  et~al., 2004, \apj, 611, 1005

\bibitem[\protect\citeauthoryear{{Gehrels}, {Sarazin}, {O'Brien}
  et~al.,}{{Gehrels} et~al.}{2005}]{gehrels05}
{Gehrels} N.,  {Sarazin} C.~L.,  {O'Brien} P.~T.,    et~al., 2005, \nat, 437,
  851

\bibitem[\protect\citeauthoryear{{Ghosh} \& {Nelemans}}{{Ghosh} \&
  {Nelemans}}{2015}]{Ghosh2015}
{Ghosh} S.,  {Nelemans} G.,  2015, Astrophysics and Space Science Proceedings,
  40, 51

\bibitem[\protect\citeauthoryear{{Giacomazzo} \& {Perna}}{{Giacomazzo} \&
  {Perna}}{2013}]{giacomazzo13}
{Giacomazzo} B.,  {Perna} R.,  2013, \apjl, 771, L26

\bibitem[\protect\citeauthoryear{{Gompertz}, {van der Horst}, {O'Brien}, {Wynn}
  \& {Wiersema}}{{Gompertz} et~al.}{2015a}]{Gompertz2015}
{Gompertz} B.~P.,  {van der Horst} A.~J.,  {O'Brien} P.~T.,  {Wynn} G.~A.,
  {Wiersema} K.,  2015a, \mnras, 448, 629

\bibitem[\protect\citeauthoryear{{Gompertz}, {van der Horst}, {O'Brien}, {Wynn}
  \& {Wiersema}}{{Gompertz} et~al.}{2015b}]{Gompertz2015MNRAS}
{Gompertz} B.~P.,  {van der Horst} A.~J.,  {O'Brien} P.~T.,  {Wynn} G.~A.,
  {Wiersema} K.,  2015b, \mnras, 448, 629

\bibitem[\protect\citeauthoryear{{G{\"o}tz} et~al.,}{{G{\"o}tz}
  et~al.}{2014}]{MXRT_2014}
{G{\"o}tz} D.,  et~al., 2014, in Society of Photo-Optical Instrumentation
  Engineers (SPIE) Conference Series Vol.~9144 of Society of Photo-Optical
  Instrumentation Engineers (SPIE) Conference Series, {The microchannel x-ray
  telescope for the gamma-ray burst mission SVOM}.
p.~23

\bibitem[\protect\citeauthoryear{{Hodapp}, {Siegmund}, {Kaiser}, {Chambers},
  {Laux}, {Morgan} \& {Mannery}}{{Hodapp} et~al.}{2004}]{Hodapp2004}
{Hodapp} K.~W.,  {Siegmund} W.~A.,  {Kaiser} N.,  {Chambers} K.~C.,  {Laux} U.,
   {Morgan} J.,    {Mannery} E.,  2004, in {Oschmann} Jr. J.~M.,  ed.,
  Ground-based Telescopes Vol.~5489 of Society of Photo-Optical Instrumentation
  Engineers (SPIE) Conference Series, {Optical design of the Pan-STARRS
  telescopes}.
pp 667--678

\bibitem[\protect\citeauthoryear{Hooper, Chung, Luan, Blair, Chen \&
  Wen}{Hooper et~al.}{2011}]{spiir}
Hooper S.,  Chung S.~K.,  Luan J.,  Blair D.,  Chen Y.,    Wen L., , 2011,
  {S}ummed {P}arallel {I}nfinite {I}mpulse {R}esponse ({S}{P}{I}{I}{R})
  {F}ilters {F}or {L}ow-{L}atency {G}ravitational {W}ave {D}etection, Phys.
  Rev. D 86, 024012 (2012)

\bibitem[\protect\citeauthoryear{{Hotokezaka}, {Kiuchi}, {Kyutoku},
  {Muranushi}, {Sekiguchi}, {Shibata} \& {Taniguchi}}{{Hotokezaka}
  et~al.}{2013}]{Hotokezaka2013}
{Hotokezaka} K.,  {Kiuchi} K.,  {Kyutoku} K.,  {Muranushi} T.,  {Sekiguchi}
  Y.-i.,  {Shibata} M.,    {Taniguchi} K.,  2013, \prd, 88, 044026

\bibitem[\protect\citeauthoryear{{Howell}, {Rowlinson}, {Coward}, {Lasky},
  {Kaplan}, {Thrane}, {Rowell}, {Galloway}, {Yuan}, {Dodson}, {Murphy}, {Hill},
  {Andreoni}, {Spitler} \& {Horton}}{{Howell} et~al.}{2015}]{Howell2015PASA}
{Howell} E.~J.,  {Rowlinson} A.,  {Coward} D.~M.,  {Lasky} P.~D.,  {Kaplan}
  D.~L.,  {Thrane} E.,  {Rowell} G.,  {Galloway} D.~K.,  {Yuan} F.,  {Dodson}
  R.,  {Murphy} T.,  {Hill} G.~C.,  {Andreoni} I.,  {Spitler} L.,    {Horton}
  A.,  2015, \pasa, 32, e046

\bibitem[\protect\citeauthoryear{Inoue et~al.,}{Inoue
  et~al.}{2013}]{Inoue2013252}
Inoue S.,  et~al., 2013, Astroparticle Physics, 43, 252

\bibitem[\protect\citeauthoryear{{Jaranowski}, {Kr{\'o}lak} \&
  {Schutz}}{{Jaranowski} et~al.}{1998}]{krolak98}
{Jaranowski} P.,  {Kr{\'o}lak} A.,    {Schutz} B.~F.,  1998, \prd, 58, 063001

\bibitem[\protect\citeauthoryear{{Johnston} et~al.,}{{Johnston}
  et~al.}{2007}]{Johnston2007PASA}
{Johnston} S.,  et~al., 2007, \pasa, 24, 174

\bibitem[\protect\citeauthoryear{{Kanner}, {Camp}, {Racusin}, {Gehrels} \&
  {White}}{{Kanner} et~al.}{2012}]{Kanner2012ApJ}
{Kanner} J.,  {Camp} J.,  {Racusin} J.,  {Gehrels} N.,    {White} D.,  2012,
  \apj, 759, 22

\bibitem[\protect\citeauthoryear{Kaplan}{Kaplan}{2015}]{Comm_MWA_2015}
Kaplan D., , 2015, personal communication

\bibitem[\protect\citeauthoryear{Kaplan et~al.,}{Kaplan
  et~al.}{2015}]{Kaplan_inprep}
Kaplan D.,  et~al., 2015, in preparation

\bibitem[\protect\citeauthoryear{{Keller}, {Schmidt}, {Bessell}, {Conroy},
  {Francis}, {Granlund}, {Kowald}, {Oates}, {Martin-Jones}, {Preston},
  {Tisserand}, {Vaccarella} \& {Waterson}}{{Keller}
  et~al.}{2007}]{Keller2007PASA}
{Keller} S.~C.,  {Schmidt} B.~P.,  {Bessell} M.~S.,  {Conroy} P.~G.,  {Francis}
  P.,  {Granlund} A.,  {Kowald} E.,  {Oates} A.~P.,  {Martin-Jones} T.,
  {Preston} T.,  {Tisserand} P.,  {Vaccarella} A.,    {Waterson} M.~F.,  2007,
  \pasa, 24, 1

\bibitem[\protect\citeauthoryear{Klimenko et~al.,}{Klimenko
  et~al.}{2011}]{klimenko}
Klimenko S.,  et~al., 2011, Phys. Rev. D, 83, 102001

\bibitem[\protect\citeauthoryear{{Klotz}, {Boer}, {Atteia}, {Gendre}, {Le
  Borgne}, {Frappa}, {Vachier} \& {Berthier}}{{Klotz}
  et~al.}{2013}]{Klotz_2013}
{Klotz} A.,  {Boer} M.,  {Atteia} J.-L.,  {Gendre} B.,  {Le Borgne} J.-F.,
  {Frappa} E.,  {Vachier} F.,    {Berthier} J.,  2013, The Messenger, 151, 6

\bibitem[\protect\citeauthoryear{{Kulkarni}}{{Kulkarni}}{2005}]{kulkarni05}
{Kulkarni} S.~R.,  2005, ArXiv Astrophysics e-prints

\bibitem[\protect\citeauthoryear{{Lasky}, {Haskell}, {Ravi}, {Howell} \&
  {Coward}}{{Lasky} et~al.}{2014}]{Lasky2014PhRvD}
{Lasky} P.~D.,  {Haskell} B.,  {Ravi} V.,  {Howell} E.~J.,    {Coward} D.~M.,
  2014, \prd, 89, 047302

\bibitem[\protect\citeauthoryear{{Lee}, {Ramirez-Ruiz} \&
  {L{\'o}pez-C{\'a}mara}}{{Lee} et~al.}{2009}]{lee09}
{Lee} W.~H.,  {Ramirez-Ruiz} E.,    {L{\'o}pez-C{\'a}mara} D.,  2009, \apjl,
  699, L93

\bibitem[\protect\citeauthoryear{{Lennarz}, {Chadwick}, {Domainko}
  et~al.,}{{Lennarz} et~al.}{2013}]{Lennarz2013}
{Lennarz} D.,  {Chadwick} P.~M.,  {Domainko} W.,    et~al., 2013,
  astro-ph:1307.6897

\bibitem[\protect\citeauthoryear{{Li} \& {Paczy{\'n}ski}}{{Li} \&
  {Paczy{\'n}ski}}{1998}]{lipaczynski98}
{Li} L.-X.,  {Paczy{\'n}ski} B.,  1998, \apjl, 507, L59

\bibitem[\protect\citeauthoryear{{LIGO Scientific Collaboration}}{{LIGO
  Scientific Collaboration}}{2009}]{SF_aLIGO}
{LIGO Scientific Collaboration} 2009, {\em Advanced LIGO anticipated
  sensitivity curves}, LIGO Document T-0900288-v2

\bibitem[\protect\citeauthoryear{Liu, Du, Chung, Hooper, Blair \& Wen}{Liu
  et~al.}{2012}]{liuyuan}
Liu Y.,  Du Z.,  Chung S.~K.,  Hooper S.,  Blair D.,    Wen L.,  2012,
  Classical and Quantum Gravity, 29, 235018

\bibitem[\protect\citeauthoryear{{Lorimer}, {Karastergiou}, {McLaughlin} \&
  {Johnston}}{{Lorimer} et~al.}{2013}]{Lorimer2013MNRAS}
{Lorimer} D.~R.,  {Karastergiou} A.,  {McLaughlin} M.~A.,    {Johnston} S.,
  2013, \mnras, 436, L5

\bibitem[\protect\citeauthoryear{{L{\"u}}, {Zhang}, {Lei}, {Li} \&
  {Lasky}}{{L{\"u}} et~al.}{2015}]{Lu2015}
{L{\"u}} H.-J.,  {Zhang} B.,  {Lei} W.-H.,  {Li} Y.,    {Lasky} P.~D.,  2015,
  ArXiv e-prints

\bibitem[\protect\citeauthoryear{{Luan}, {Hooper}, {Wen} \& {Chen}}{{Luan}
  et~al.}{2011}]{jing}
{Luan} J.,  {Hooper} S.,  {Wen} L.,    {Chen} Y.,  2011, ArXiv e-prints
  1108.3174

\bibitem[\protect\citeauthoryear{{Manzotti} \& {Dietz}}{{Manzotti} \&
  {Dietz}}{2012}]{manzotti12}
{Manzotti} A.,  {Dietz} A.,  2012, ArXiv e-prints

\bibitem[\protect\citeauthoryear{Meegan et~al.,}{Meegan
  et~al.}{2009}]{Meegan_Fermi_2009}
Meegan C.,  et~al., 2009, \apj, 702, 791

\bibitem[\protect\citeauthoryear{{Meszaros} \& {Rees}}{{Meszaros} \&
  {Rees}}{1993}]{Meszaros_Rees1993ApJ}
{Meszaros} P.,  {Rees} M.~J.,  1993, \apj, 405, 278

\bibitem[\protect\citeauthoryear{{Metzger}, {Bauswein}, {Goriely} \&
  {Kasen}}{{Metzger} et~al.}{2015}]{Metzger_2014}
{Metzger} B.~D.,  {Bauswein} A.,  {Goriely} S.,    {Kasen} D.,  2015, \mnras,
  446, 1115

\bibitem[\protect\citeauthoryear{{Metzger} \& {Berger}}{{Metzger} \&
  {Berger}}{2012}]{MetzgerBerger2012ApJ}
{Metzger} B.~D.,  {Berger} E.,  2012, \apj, 746, 48

\bibitem[\protect\citeauthoryear{{Metzger} \& {Bower}}{{Metzger} \&
  {Bower}}{2014}]{Metzger2014b}
{Metzger} B.~D.,  {Bower} G.~C.,  2014, \mnras, 437, 1821

\bibitem[\protect\citeauthoryear{{Metzger}, {Giannios}, {Thompson},
  {Bucciantini} \& {Quataert}}{{Metzger} et~al.}{2011}]{Metzger2011}
{Metzger} B.~D.,  {Giannios} D.,  {Thompson} T.~A.,  {Bucciantini} N.,
  {Quataert} E.,  2011, \mnras, 413, 2031

\bibitem[\protect\citeauthoryear{{Metzger}, {Mart{\'{\i}}nez-Pinedo}, {Darbha},
  {Quataert}, {Arcones}, {Kasen}, {Thomas}, {Nugent}, {Panov} \&
  {Zinner}}{{Metzger} et~al.}{2010}]{metzger10}
{Metzger} B.~D.,  {Mart{\'{\i}}nez-Pinedo} G.,  {Darbha} S.,  {Quataert} E.,
  {Arcones} A.,  {Kasen} D.,  {Thomas} R.,  {Nugent} P.,  {Panov} I.~V.,
  {Zinner} N.~T.,  2010, \mnras, 406, 2650

\bibitem[\protect\citeauthoryear{{Metzger} \& {Piro}}{{Metzger} \&
  {Piro}}{2014}]{metzgerpiro14}
{Metzger} B.~D.,  {Piro} A.~L.,  2014, \mnras, 439, 3916

\bibitem[\protect\citeauthoryear{{Metzger}, {Quataert} \& {Thompson}}{{Metzger}
  et~al.}{2008}]{metzger08}
{Metzger} B.~D.,  {Quataert} E.,    {Thompson} T.~A.,  2008, \mnras, 385, 1455

\bibitem[\protect\citeauthoryear{{Murphy} et~al.,}{{Murphy}
  et~al.}{2013}]{Murphy2013PASAa}
{Murphy} T.,  et~al., 2013, \pasa, 30, 6

\bibitem[\protect\citeauthoryear{{Nakar} \& {Piran}}{{Nakar} \&
  {Piran}}{2011}]{nakar11}
{Nakar} E.,  {Piran} T.,  2011, \nat, 478, 82

\bibitem[\protect\citeauthoryear{{Narayan}, {Paczynski} \& {Piran}}{{Narayan}
  et~al.}{1992a}]{Narayan_1992ApJ}
{Narayan} R.,  {Paczynski} B.,    {Piran} T.,  1992a, \apjl, 395, L83

\bibitem[\protect\citeauthoryear{{Narayan}, {Paczynski} \& {Piran}}{{Narayan}
  et~al.}{1992b}]{narayan92}
{Narayan} R.,  {Paczynski} B.,    {Piran} T.,  1992b, \apjl, 395, L83

\bibitem[\protect\citeauthoryear{Nissanke, Holz, Hughes, Dalal \&
  Sievers}{Nissanke et~al.}{2009}]{nissanke09}
Nissanke S.,  Holz D.~E.,  Hughes S.~A.,  Dalal N.,    Sievers J.~L., , 2009,
  {E}xploring short gamma-ray bursts as gravitational-wave standard sirens,
  Astrophys.J.725:496-514,2010

\bibitem[\protect\citeauthoryear{Nissanke, Kasliwal \& Georgieva}{Nissanke
  et~al.}{2012}]{nissanke12}
Nissanke S.,  Kasliwal M.,    Georgieva A., , 2012, Astrophysical Journal, 767,
  124 (2013)

\bibitem[\protect\citeauthoryear{{Nissanke}, {Kasliwal} \&
  {Georgieva}}{{Nissanke} et~al.}{2013}]{Nissanke2013ApJ}
{Nissanke} S.,  {Kasliwal} M.,    {Georgieva} A.,  2013, \apj, 767, 124

\bibitem[\protect\citeauthoryear{Nissanke, Sievers, Dalal \& Holz}{Nissanke
  et~al.}{2011}]{samaya}
Nissanke S.,  Sievers J.,  Dalal N.,    Holz D.,  2011, Astrophys. J., 739, 99

\bibitem[\protect\citeauthoryear{{Norris} \& {Bonnell}}{{Norris} \&
  {Bonnell}}{2006}]{norris06}
{Norris} J.~P.,  {Bonnell} J.~T.,  2006, \apj, 643, 266

\bibitem[\protect\citeauthoryear{O'Brien \& Hinton}{O'Brien \&
  Hinton}{2015}]{Comm_CTA_2015}
O'Brien P.,  Hinton J., , 2015, personal communication

\bibitem[\protect\citeauthoryear{{Oosterloo}, {Verheijen} \& {van
  Cappellen}}{{Oosterloo} et~al.}{2010}]{Oosterloo2010}
{Oosterloo} T.,  {Verheijen} M.,    {van Cappellen} W.,  2010, in ISKAF2010
  Science Meeting {The latest on Apertif}.
p.~43

\bibitem[\protect\citeauthoryear{{\"O}zel, Psaltis, Narayan \&
  Villarreal}{{\"O}zel et~al.}{2012}]{ozel2012mass}
{\"O}zel F.,  Psaltis D.,  Narayan R.,    Villarreal A.~S.,  2012, The
  Astrophysical Journal, 757, 55

\bibitem[\protect\citeauthoryear{{Paczynski}}{{Paczynski}}{1986}]{Paczynski1986ApJ}
{Paczynski} B.,  1986, \apjl, 308, L43

\bibitem[\protect\citeauthoryear{{Pacz\'ynski}}{{Pacz\'ynski}}{1986}]{paczynski86}
{Pacz\'ynski} B.,  1986, \apjl, 308, L43

\bibitem[\protect\citeauthoryear{{Pacz\'ynski}}{{Pacz\'ynski}}{1991}]{paczynski91}
{Pacz\'ynski} B.,  1991, Acta Astronomica, 41, 257

\bibitem[\protect\citeauthoryear{{Page}, {Willingale}, {Osborne}
  et~al.,}{{Page} et~al.}{2007}]{page07}
{Page} K.~L.,  {Willingale} R.,  {Osborne} J.~P.,    et~al., 2007, \apj, 663,
  1125

\bibitem[\protect\citeauthoryear{{Paul}, {Wei}, {Basa} \& {Zhang}}{{Paul}
  et~al.}{2011}]{Paul2011CRPhy}
{Paul} J.,  {Wei} J.,  {Basa} S.,    {Zhang} S.-N.,  2011, Comptes Rendus
  Physique, 12, 298

\bibitem[\protect\citeauthoryear{{Perley}, {Cenko}, {Bloom}, {Chen}, {Butler},
  {Kocevski}, {Prochaska}, {Brodwin}, {Glazebrook}, {Kasliwal}, {Kulkarni},
  {Lopez}, {Ofek}, {Pettini}, {Soderberg} \& {Starr}}{{Perley}
  et~al.}{2009}]{perley09}
{Perley} D.~A.,  {Cenko} S.~B.,  {Bloom} J.~S.,  {Chen} H.-W.,  {Butler} N.~R.,
   {Kocevski} D.,  {Prochaska} J.~X.,  {Brodwin} M.,  {Glazebrook} K.,
  {Kasliwal} M.~M.,  {Kulkarni} S.~R.,  {Lopez} S.,  {Ofek} E.~O.,  {Pettini}
  M.,  {Soderberg} A.~M.,    {Starr} D.,  2009, \aj, 138, 1690

\bibitem[\protect\citeauthoryear{{Perley}, {Cenko}, {Corsi} et~al.,}{{Perley}
  et~al.}{2014}]{Perley2014ApJ}
{Perley} D.~A.,  {Cenko} S.~B.,  {Corsi} A.,    et~al., 2014, \apj, 781, 37

\bibitem[\protect\citeauthoryear{{Piran}, {Nakar} \& {Rosswog}}{{Piran}
  et~al.}{2013}]{piran13}
{Piran} T.,  {Nakar} E.,    {Rosswog} S.,  2013, \mnras, 430, 2121

\bibitem[\protect\citeauthoryear{{Prasad}, {Wijnholds}, {Huizinga} \&
  {Wijers}}{{Prasad} et~al.}{2014}]{Prasad2014}
{Prasad} P.,  {Wijnholds} S.~J.,  {Huizinga} F.,    {Wijers} R.~A.~M.~J.,
  2014, \aap, 568, A48

\bibitem[\protect\citeauthoryear{{Punturo}}{{Punturo}}{2012}]{SFaVirgo}
{Punturo} M.,  2012, note vir-0128a-12, Advanced Virgo Technical Design Report.
The Virgo Collaboration

\bibitem[\protect\citeauthoryear{{Racusin}, {Karpov}, {Sokolowski}
  et~al.,}{{Racusin} et~al.}{2008}]{Racusin2008Natur}
{Racusin} J.~L.,  {Karpov} S.~V.,  {Sokolowski} M.,    et~al., 2008, \nat, 455,
  183

\bibitem[\protect\citeauthoryear{{Ravi} \& {Lasky}}{{Ravi} \&
  {Lasky}}{2014}]{Ravi2014MNRAS}
{Ravi} V.,  {Lasky} P.~D.,  2014, \mnras, 441, 2433

\bibitem[\protect\citeauthoryear{{Regimbau}, {Siellez}, {Meacher}, {Gendre} \&
  {Bo{\"e}r}}{{Regimbau} et~al.}{2015}]{Regimbau2015ApJ}
{Regimbau} T.,  {Siellez} K.,  {Meacher} D.,  {Gendre} B.,    {Bo{\"e}r} M.,
  2015, \apj, 799, 69

\bibitem[\protect\citeauthoryear{{Rezzolla}, {Giacomazzo}, {Baiotti}, {Granot},
  {Kouveliotou} \& {Aloy}}{{Rezzolla} et~al.}{2011}]{Rezzolla_2011ApJ}
{Rezzolla} L.,  {Giacomazzo} B.,  {Baiotti} L.,  {Granot} J.,  {Kouveliotou}
  C.,    {Aloy} M.~A.,  2011, \apjl, 732, L6

\bibitem[\protect\citeauthoryear{{Rezzolla} \& {Kumar}}{{Rezzolla} \&
  {Kumar}}{2014}]{Rezzolla2014}
{Rezzolla} L.,  {Kumar} P.,  2014, ArXiv e-prints

\bibitem[\protect\citeauthoryear{{Rosswog}, {Piran} \& {Nakar}}{{Rosswog}
  et~al.}{2013}]{rosswog13}
{Rosswog} S.,  {Piran} T.,    {Nakar} E.,  2013, \mnras, 430, 2585

\bibitem[\protect\citeauthoryear{{Rowlinson} et~al.,}{{Rowlinson}
  et~al.}{2010}]{Rowlinson2010MNRAS}
{Rowlinson} A.,  et~al., 2010, \mnras, 409, 531

\bibitem[\protect\citeauthoryear{{Rowlinson}, {O'Brien}, {Metzger}, {Tanvir} \&
  {Levan}}{{Rowlinson} et~al.}{2013}]{Rowlinson2013MNRAS}
{Rowlinson} A.,  {O'Brien} P.~T.,  {Metzger} B.~D.,  {Tanvir} N.~R.,    {Levan}
  A.~J.,  2013, \mnras, 430, 1061

\bibitem[\protect\citeauthoryear{{Rowlinson}, {Wiersema}, {Levan}, {Tanvir},
  {O'Brien}, {Rol}, {Hjorth}, {Th{\"o}ne}, {de Ugarte Postigo}, {Fynbo},
  {Jakobsson}, {Pagani} \& {Stamatikos}}{{Rowlinson}
  et~al.}{2010a}]{Rowlinson2010}
{Rowlinson} A.,  {Wiersema} K.,  {Levan} A.~J.,  {Tanvir} N.~R.,  {O'Brien}
  P.~T.,  {Rol} E.,  {Hjorth} J.,  {Th{\"o}ne} C.~C.,  {de Ugarte Postigo} A.,
  {Fynbo} J.~P.~U.,  {Jakobsson} P.,  {Pagani} C.,    {Stamatikos} M.,  2010a,
  \mnras, 408, 383

\bibitem[\protect\citeauthoryear{{Rowlinson}, {Wiersema}, {Levan}, {Tanvir},
  {O'Brien}, {Rol}, {Hjorth}, {Th{\"o}ne}, {de Ugarte Postigo}, {Fynbo},
  {Jakobsson}, {Pagani} \& {Stamatikos}}{{Rowlinson}
  et~al.}{2010b}]{Rowlinson2010MNRASb}
{Rowlinson} A.,  {Wiersema} K.,  {Levan} A.~J.,  {Tanvir} N.~R.,  {O'Brien}
  P.~T.,  {Rol} E.,  {Hjorth} J.,  {Th{\"o}ne} C.~C.,  {de Ugarte Postigo} A.,
  {Fynbo} J.~P.~U.,  {Jakobsson} P.,  {Pagani} C.,    {Stamatikos} M.,  2010b,
  \mnras, 408, 383

\bibitem[\protect\citeauthoryear{Schutz}{Schutz}{2011}]{schutz11}
Schutz B.~F.,  2011, Classical and Quantum Gravity, 28, 125023

\bibitem[\protect\citeauthoryear{{Shawhan}}{{Shawhan}}{2012}]{Shawhan2012SPIE}
{Shawhan} P.~S.,  2012, in Society of Photo-Optical Instrumentation Engineers
  (SPIE) Conference Series Vol.~8448 of Society of Photo-Optical
  Instrumentation Engineers (SPIE) Conference Series, {Rapid alerts for
  following up gravitational wave event candidates}.
p.~0

\bibitem[\protect\citeauthoryear{{Shibata}, {Taniguchi}, {Okawa} \&
  {Buonanno}}{{Shibata} et~al.}{2014}]{Shibata2014}
{Shibata} M.,  {Taniguchi} K.,  {Okawa} H.,    {Buonanno} A.,  2014, \prd, 89,
  084005

\bibitem[\protect\citeauthoryear{{Siellez}, {Bo{\"e}r} \& {Gendre}}{{Siellez}
  et~al.}{2014}]{Siellez2014MNRAS}
{Siellez} K.,  {Bo{\"e}r} M.,    {Gendre} B.,  2014, \mnras, 437, 649

\bibitem[\protect\citeauthoryear{Singer \& Price}{Singer \&
  Price}{2015}]{leo_bayestar}
Singer L.,  Price L.,  2015, in preparation

\bibitem[\protect\citeauthoryear{{Singer}}{{Singer}}{2015}]{Singer_2015PhDT}
{Singer} L.~P.,  2015, PhD thesis, California Institute of Technology

\bibitem[\protect\citeauthoryear{{Singer}, {Cenko}, {Kasliwal}
  et~al.,}{{Singer} et~al.}{2013}]{Singer2013}
{Singer} L.~P.,  {Cenko} S.~B.,  {Kasliwal} M.~M.,    et~al., 2013, \apjl, 776,
  L34

\bibitem[\protect\citeauthoryear{{Singer}, {Kasliwal}, {Cenko}
  et~al.,}{{Singer} et~al.}{2015}]{SingerPTF2015ApJ}
{Singer} L.~P.,  {Kasliwal} M.~M.,  {Cenko} S.~B.,    et~al., 2015, \apj, 806,
  52

\bibitem[\protect\citeauthoryear{{Singer}, {Price}, {Farr} et~al.,}{{Singer}
  et~al.}{2014}]{first2years}
{Singer} L.~P.,  {Price} L.~R.,  {Farr} B.,    et~al., 2014, \apj, 795, 105

\bibitem[\protect\citeauthoryear{{Smith}, {Dekany}, {Bebek}, {Bellm}, {Bui},
  {Cromer}, {Gardner}, {Hoff}, {Kaye}, {Kulkarni}, {Lambert}, {Levi} \&
  {Reiley}}{{Smith} et~al.}{2014}]{Smith_2014SPIE}
{Smith} R.~M.,  {Dekany} R.~G.,  {Bebek} C.,  {Bellm} E.,  {Bui} K.,  {Cromer}
  J.,  {Gardner} P.,  {Hoff} M.,  {Kaye} S.,  {Kulkarni} S.,  {Lambert} A.,
  {Levi} M.,    {Reiley} D.,  2014, in Society of Photo-Optical Instrumentation
  Engineers (SPIE) Conference Series Vol.~9147 of Society of Photo-Optical
  Instrumentation Engineers (SPIE) Conference Series, {The Zwicky transient
  facility observing system}.
p.~79

\bibitem[\protect\citeauthoryear{Somiya}{Somiya}{2012}]{kagra_2012}
Somiya K.,  2012, Classical and Quantum Gravity, 29, 124007

\bibitem[\protect\citeauthoryear{{Staley}, {Titterington}, {Fender},
  {Swinbank}, {van der Horst}, {Rowlinson}, {Scaife}, {Grainge} \&
  {Pooley}}{{Staley} et~al.}{2013}]{Staley2013MNRAS}
{Staley} T.~D.,  {Titterington} D.~J.,  {Fender} R.~P.,  {Swinbank} J.~D.,
  {van der Horst} A.~J.,  {Rowlinson} A.,  {Scaife} A.~M.~M.,  {Grainge}
  K.~J.~B.,    {Pooley} G.~G.,  2013, \mnras, 428, 3114

\bibitem[\protect\citeauthoryear{Steeghs \& Galloway}{Steeghs \&
  Galloway}{2015}]{Comm_GOTO_2015}
Steeghs D. T.~H.,  Galloway D.~K., , 2015, personal communication

\bibitem[\protect\citeauthoryear{{Tanaka} \& {Hotokezaka}}{{Tanaka} \&
  {Hotokezaka}}{2013}]{Tanaka2013}
{Tanaka} M.,  {Hotokezaka} K.,  2013, \apj, 775, 113

\bibitem[\protect\citeauthoryear{{Tanvir}, {Levan}, {Fruchter}, {Hjorth},
  {Hounsell}, {Wiersema} \& {Tunnicliffe}}{{Tanvir}
  et~al.}{2013}]{Tanvir2013Natur}
{Tanvir} N.~R.,  {Levan} A.~J.,  {Fruchter} A.~S.,  {Hjorth} J.,  {Hounsell}
  R.~A.,  {Wiersema} K.,    {Tunnicliffe} R.~L.,  2013, \nat, 500, 547

\bibitem[\protect\citeauthoryear{{Thornton} et~al.,}{{Thornton}
  et~al.}{2013}]{Thornton2013Sci}
{Thornton} D.,  et~al., 2013, Science, 341, 53

\bibitem[\protect\citeauthoryear{{Tingay} et~al.,}{{Tingay}
  et~al.}{2013}]{Tingay2013PASA}
{Tingay} S.~J.,  et~al., 2013, \pasa, 30, 7

\bibitem[\protect\citeauthoryear{{Totani}}{{Totani}}{2013}]{Totani2013PASJ}
{Totani} T.,  2013, \pasj, 65, L12

\bibitem[\protect\citeauthoryear{{Tremblay}, {Ord}, {Bhat} et~al.,}{{Tremblay}
  et~al.}{2015}]{Tremblay2015}
{Tremblay} S.~E.,  {Ord} S.~M.,  {Bhat} N.~D.~R.,    et~al., 2015, \pasa, 32, 5

\bibitem[\protect\citeauthoryear{{Troja}, {Cusumano}, {O'Brien}
  et~al.,}{{Troja} et~al.}{2007}]{Troja07}
{Troja} E.,  {Cusumano} G.,  {O'Brien} P.~T.,    et~al., 2007, \apj, 665, 599

\bibitem[\protect\citeauthoryear{{Tunnicliffe}, {Levan}, {Tanvir}, {Rowlinson},
  {Perley}, {Bloom}, {Cenko}, {O'Brien}, {Cobb}, {Wiersema}, {Malesani}, {de
  Ugarte Postigo}, {Hjorth}, {Fynbo} \& {Jakobsson}}{{Tunnicliffe}
  et~al.}{2014}]{Tunnicliffe2014MNRAS}
{Tunnicliffe} R.~L.,  {Levan} A.~J.,  {Tanvir} N.~R.,  {Rowlinson} A.,
  {Perley} D.~A.,  {Bloom} J.~S.,  {Cenko} S.~B.,  {O'Brien} P.~T.,  {Cobb}
  B.~E.,  {Wiersema} K.,  {Malesani} D.,  {de Ugarte Postigo} A.,  {Hjorth} J.,
   {Fynbo} J.~P.~U.,    {Jakobsson} P.,  2014, \mnras, 437, 1495

\bibitem[\protect\citeauthoryear{{Usov} \& {Katz}}{{Usov} \&
  {Katz}}{2000}]{Usov2000A&A}
{Usov} V.~V.,  {Katz} J.~I.,  2000, \aap, 364, 655

\bibitem[\protect\citeauthoryear{{van der Horst}, {Paragi}, {de Bruyn}
  et~al.,}{{van der Horst} et~al.}{2014}]{vanderhorst2014}
{van der Horst} A.~J.,  {Paragi} Z.,  {de Bruyn} A.~G.,    et~al., 2014,
  \mnras, 444, 3151

\bibitem[\protect\citeauthoryear{{van Haarlem}, {Wise}, {Gunst} et~al.,}{{van
  Haarlem} et~al.}{2013}]{vanHaarlem2013AA}
{van Haarlem} M.~P.,  {Wise} M.~W.,  {Gunst} A.~W.,    et~al., 2013, \aap, 556,
  A2

\bibitem[\protect\citeauthoryear{{Veres} \& {M{\'e}sz{\'a}ros}}{{Veres} \&
  {M{\'e}sz{\'a}ros}}{2014}]{Veres_2014ApJ}
{Veres} P.,  {M{\'e}sz{\'a}ros} P.,  2014, \apj, 787, 168

\bibitem[\protect\citeauthoryear{{Verheijen}, {Oosterloo}, {van Cappellen},
  {Bakker}, {Ivashina} \& {van der Hulst}}{{Verheijen}
  et~al.}{2008}]{Verheijen2008}
{Verheijen} M.~A.~W.,  {Oosterloo} T.~A.,  {van Cappellen} W.~A.,  {Bakker} L.,
   {Ivashina} M.~V.,    {van der Hulst} J.~M.,  2008, in {Minchin} R.,
  {Momjian} E.,  eds, The Evolution of Galaxies Through the Neutral Hydrogen
  Window Vol.~1035 of American Institute of Physics Conference Series,
  {Apertif, a focal plane array for the WSRT}.
pp 265--271

\bibitem[\protect\citeauthoryear{{Vestrand} et~al.,}{{Vestrand}
  et~al.}{2005}]{Vestrand2005Natur}
{Vestrand} W.~T.,  et~al., 2005, \nat, 435, 178

\bibitem[\protect\citeauthoryear{{Vestrand}, {Wren}, {Panaitescu}, {Wozniak},
  {Davis}, {Palmer}, {Vianello}, {Omodei}, {Xiong}, {Briggs}, {Elphick},
  {Paciesas} \& {Rosing}}{{Vestrand} et~al.}{2014}]{Vestrand2014Sci}
{Vestrand} W.~T.,  {Wren} J.~A.,  {Panaitescu} A.,  {Wozniak} P.~R.,  {Davis}
  H.,  {Palmer} D.~M.,  {Vianello} G.,  {Omodei} N.,  {Xiong} S.,  {Briggs}
  M.~S.,  {Elphick} M.,  {Paciesas} W.,    {Rosing} W.,  2014, Science, 343, 38

\bibitem[\protect\citeauthoryear{{Vitale} \& {Zanolin}}{{Vitale} \&
  {Zanolin}}{2011}]{frequentist_network}
{Vitale} S.,  {Zanolin} M.,  2011, \prd, 84, 104020

\bibitem[\protect\citeauthoryear{Wen \& Chen}{Wen \& Chen}{2010}]{wen10}
Wen L.,  Chen Y.,  2010, Phys. Rev. D, 81, 082001

\bibitem[\protect\citeauthoryear{{Wen}, {Fan} \& {Chen}}{{Wen}
  et~al.}{2008}]{wen_fan}
{Wen} L.,  {Fan} X.,    {Chen} Y.,  2008, Journal of Physics Conference Series,
  122, 012038

\bibitem[\protect\citeauthoryear{{Yang}, {Jin}, {Li}, {Covino}, {Zheng},
  {Hotokezaka}, {Fan}, {Piran} \& {Wei}}{{Yang} et~al.}{2015}]{yang15}
{Yang} B.,  {Jin} Z.-P.,  {Li} X.,  {Covino} S.,  {Zheng} X.-Z.,  {Hotokezaka}
  K.,  {Fan} Y.-Z.,  {Piran} T.,    {Wei} D.-M.,  2015, Nature Communications,
  6, 7323

\bibitem[\protect\citeauthoryear{{Yu} \& {Huang}}{{Yu} \&
  {Huang}}{2007}]{Yu2007}
{Yu} Y.,  {Huang} Y.-F.,  2007, \cjaa, 7, 669

\bibitem[\protect\citeauthoryear{{Yu}, {Zhang} \& {Gao}}{{Yu}
  et~al.}{2013}]{yu13}
{Yu} Y.-W.,  {Zhang} B.,    {Gao} H.,  2013, \apjl, 776, L40

\bibitem[\protect\citeauthoryear{Zanolin, Vitale \& Makris}{Zanolin
  et~al.}{2009}]{frequentist_single}
Zanolin M.,  Vitale S.,    Makris N., , 2009, Phys.Rev.D81:124048,2010

\bibitem[\protect\citeauthoryear{{Zhang}}{{Zhang}}{2013}]{Zhang2013ApJ}
{Zhang} B.,  2013, \apjl, 763, L22

\bibitem[\protect\citeauthoryear{{Zhang}}{{Zhang}}{2014}]{Zhang2014ApJ}
{Zhang} B.,  2014, \apjl, 780, L21

\bibitem[\protect\citeauthoryear{{Zhang}, {Kobayashi} \&
  {M{\'e}sz{\'a}ros}}{{Zhang} et~al.}{2003}]{Zhang03_ApJ}
{Zhang} B.,  {Kobayashi} S.,    {M{\'e}sz{\'a}ros} P.,  2003, \apj, 595, 950

\bibitem[\protect\citeauthoryear{{Zhang} \& {M{\'e}sz{\'a}ros}}{{Zhang} \&
  {M{\'e}sz{\'a}ros}}{2001}]{Zhang2001}
{Zhang} B.,  {M{\'e}sz{\'a}ros} P.,  2001, \apjl, 552, L35

\bibitem[\protect\citeauthoryear{{Zheng}, {Shen}, {Sakamoto}, {Beardmore}, {De
  Pasquale}, {Wu}, {Gorosabel}, {Urata}, {Sugita} \& {Zhang}}{{Zheng}
  et~al.}{2012}]{zheng12}
{Zheng} W.,  {Shen} R.~F.,  {Sakamoto} T.,  {Beardmore} A.~P.,  {De Pasquale}
  M.,  {Wu} X.~F.,  {Gorosabel} J.,  {Urata} Y.,  {Sugita} S.,    {Zhang} B.
  e.~a.,  2012, \apj, 751, 90

\end{thebibliography}

 \label{lastpage}
\end{document}